\documentclass[12pt]{article}

\usepackage{amsmath}
\usepackage{amssymb}
\usepackage{amsfonts}
\usepackage{graphicx}
\numberwithin{equation}{section}

\begin{document}

\begin{center}
{\large \bf{ P-stars in the  gravitational wave era  }}
\end{center}

\vspace*{1.0 cm}

\begin{center}
{
Paolo Cea~\protect\footnote{Electronic address:
{\tt paolo.cea@ba.infn.it}}  \\[0.5cm]
{\em INFN - Sezione di Bari, Via Amendola 173 - 70126 Bari,
Italy} }
\end{center}

\vspace*{1.0 cm}

\begin{abstract}
\noindent 
P-stars are compact relativistic stars made of deconfined up and down quarks in a chromomagnetic condensate proposed
by us long time ago. P-stars do not admit a critical mass thereby they are able to overcome the gravitational collapse
to black holes. In this work we discuss in greater details our theoretical proposal for P-stars.  We point out that our theory
for compact relativistic stars stems from  our own understanding of the confining  quantum vacuum 
supported by estensive non-perturbative numerical simulations of Quantum ChromoDynamics on the lattice.
We compare our proposal with the constraints arising from the recent observations of massive pulsars,
the gravitational event GW170817 and the precise determination of  the PSR J0030+0451 mass and radius from NICER data.
We argue that core-collapsed supernovae could give rise to a P-star instead of a neutron star. In this case we show that
the birth of a P-star could solve the supernova explosion problem leading to successful supernova explosions with total
energies up to $10^{53}$ erg.
We critically compare P-stars with the gravitational wave event GW170817 and the subsequent electromagnetic follow-up,
the  short Gamma Ray Burst GRB170817A and  the kilonova  AT2017gfo.
We  also present an explorative study on gravitational wave emission from coalescing binary P-stars with masses 
$M_1 \simeq M_2 \simeq 30 \; M_{\odot}$. We attempt a qualitative comparison with the gravitational wave event GW150914. 
We find that the gravitational wave strain amplitude  from massive P-star binaries could mimic the ringdown gravitational wave emission 
by coalescing black hole binaries.  We point out that a clear signature for massive P-stars would be the detection of  wobble
frequencies in the gravitational wave strain amplitude in the post-merger phase of two coalescing massive compact objects with unequal masses.
\end{abstract}

\vspace*{0.6cm}

\vspace{0.2cm}
\newpage
\section{Introduction}
\label{S1}
The detections in the first and second runs by the Advanced  Laser  Interferometer Gravitational -wave Observatory (LIGO)
 Collaborations~\cite{LIGO:2016a,LIGO:2016b,LIGO:2016c,LIGO:2017a,LIGO:2017b} and by the LIGO and 
 Virgo Collaborations~\cite{LIGO:2017c,LIGO:2017d}  of
gravitational wave signals from binary compact object mergers (see, also, the LIGO and Virgo gravitational wave transient
catalog~\cite{LIGO-Virgo-catalog:2018}) have opened up the possibility of new tests on the true nature of 
these objects~\cite{LIGO:2016d,LIGO:2016e}. It turned out that the involved objects were extremely small~\cite{LIGO:2017e}.
Moreover, almost all the gravitational wave events during the first and the second observing runs 
involved compact objects with mass well above  $3 \;  M_{\odot}$,   $M_{\odot}$
being the solar mass. The unique exception  was the gravitational wave signal denoted GW170817, 
observed by the LIGO and Virgo Collaborations on 17 August  2017~\cite{LIGO:2017d} 
that involved  inspiraling  objects with masses in the range of $1.1$ to 
$1.6 \;  M_{\odot}$. Follow-up observations by other telescopes~\cite{LIGO:2017f,LIGO:2017g} revealed emissions at various wavelengths 
opening, thus, the window of the multi-messenger astronomy era. \\
The most compact stars known are pulsars. Actually,  soon after  the discovery of the first pulsar~\cite{Hewish:1968}, it becomes generally accepted
that the pulsars are rapidly rotating neutron stars~\cite{Baade:1934a,Baade:1934b} endowed with a strong magnetic 
field~\cite{Gold:1968,Pacini:1968}. 
Considering that in the  almost all gravitational wave events detected up to now during the first two LIGO-Virgo runs
the involved masses of the compact objects are well above the neutron star mass limit, it is widely believed that these compact bodies are black holes.
This conclusion arises from the almost  unanimous conviction  that there are no alternative models able to provide as satisfactory
 an explanation for the wide variety of pulsar  phenomena as those built around the rotating neutron star model together with
 the existence of a critical mass of about $3 \,  M_{\odot}$ above which a neutron star
cannot be maintained against gravity leading to the gravitational collapse to black hole.
In addition, the neutron star model seems to gaining further support from the gravitational wave event GW170817 triggered by inspiraling 
solar-mass compact objects due the observation~\cite{LIGO:2017f,LIGO:2017g} of the expected short Gamma-Ray-Burst event GRB170817A
delayed by about $1.7 \; $s with respect to the merger time, and the optical counterpart, called kilonova AT2017gfo, because their peak luminosity
were consistent with the expectations that neutron star mergers host r-processes responsible of the synthesis of the most heavy nuclei. \\
 Nevertheless, several years ago  we advanced the proposal for  a new class  of compact stars, named P-stars, that challenged  the standard  
 paradigm based on neutron stars and black holes (see, for instance Ref.~\cite{Shapiro:1983}). P-stars are compact relativistic stars made 
 of deconfined up and down quarks in $\beta$-equilibrium with electrons in a chromomagnetic  condensate.
P-stars do not admit a limit mass thereby they are able to overcome the gravitational collapse to black hole. 
The basis of the physical model of P-stars has been presented in a series of papers~\cite{Cea:2004a,Cea:2004b,Cea:2006,Cea:2008}.
We found that P-stars were able to account for relativistic compact stars, namely stars with masses and radii comparable with canonical
neutron stars, as well as supermassive compact objects. Moreover, we suggested that P-stars once formed were absolutely stable.
This led us to advance the proposal that pulsars could be P-stars instead of neutron stars. Obviously, this drastic paradigm shift required
to ascertain if P-stars are able to account for the huge observed phenomenology of pulsars. In fact, in our previous 
papers~\cite{Cea:2004a,Cea:2004b,Cea:2006,Cea:2008} we found that the dominant cooling processes in P-stars were by neutrino emission
via direct $\beta$-decay quark reactions (the so-called URCA processes~\cite{Shapiro:1983}). Moreover, it turned out that the resulting
cooling curves of P-stars compared in a satisfying way with several observations. We argued that our model could easily accomodate
the observations from nearest isolated compact stars. In addition, we discussed P-stars endowed with super strong dipolar magnetic fields.
We found that soft gamma-ray repeaters and anomalous X-ray pulsars could be understood within our model. In particular, we showed
that there is a quite natural mechanism for the generation of almost dipolar surface magnetic fields up to $10^{16} \, $G.
We argued that in P-stars the glitches are triggered by magnetic dissipative effects in the inner core. Moreover, we found that for strong magnetic
fields these magnetic glitches are at the origin of both the quiescent emissions and burst activities. As a consequence, we are confident that
our results indicated that P-stars are able to uncover several observational features from isolated and binary pulsars. \\
It is not an exaggeration to say that one of the most dramatic prediction of our theory is that there may exist compact stars with masses well
above the neutron star limiting mass $M \, \gtrsim  \, 3 \,  M_{\odot}$. Within the standard paradigm these objects are interpreted as black holes.
However, from the observational point of view it is not easy to distinguish between a true black hole and a massive compact object
with radius slightly larger than the Schwarzschild radius $R_S \, = \, 2 \, G \, M /  c^2$ (see, eg, the recent review Ref.~\cite{Cardoso:2019}
and references therein).
More recently, the Event Horizon Telescope Collaboration has succeeded in imaging the so-called shadow of a central massive object in the elliptical
galaxy M87 as an evidence for the existence of a black hole with several billion of solar
 masses~\cite{EHT:2019a,EHT:2019b,EHT:2019c,EHT:2019d,EHT:2019e,EHT:2019f}. 
Nevertheless, a sufficiently compact and
rotating object endowed with a strong enough magnetic field cannot be yet excluded. In our opinion the most compelling evidence for black holes
comes from the observations of gravitational waves emitted during the ringdown phase when two massive compact objects coalesce and the
merger remnant settles down to stationary equilibrium. Actually, the resulting gravitational wave strain amplitude in the ringdown phase seem to
be described very well by strong-field perturbative techniques if one assumes that  the  coalescing compact objects are black holes. \\
With this paper we aim to  critically compare P-stars with recent advances in astrophysical observations. In particular we will
discuss the constraints arising from the recent detections of pulsars with mass exceeding two solar masses. In addition, we will take into account
the limits  set by the gravitational event GW170817  on the equation of state of nuclear matter at high densities.  
We will, also,  present explorative calculations aimed at illustrate as simple as possible the physical mechanisms responsible
for the gravitational wave emission from binary coalescing P-stars. We will focus on massive P-stars with masses  
$M_1 \, \simeq \, M_2 \, \simeq  \, 30 \,  M_{\odot}$ and attempt a qualitative comparison with  the first LIGO gravitational event GW150914. 
Indeed, we find that the merger of binaries containing massive P-stars can mimic the characteristic quasinormal ringing signal expected in the  gravitational wave strain amplitude of two black holes in the ringdown phase. \\
The plan of the paper as following.
In Sect.~\ref{S2}  we discuss in greater details our theoretical proposal for P-stars. In particular, 
an effort is made to show how  our theory for compact relativistic stars stems from  our own understanding 
of the confining  quantum vacuum  supported by estensive non-perturbative numerical simulations of
Quantum Chromodynamics (QCD) on the lattice. Moreover,  we suggest that core-collapsed supernovae could give rise 
to a P-star instead of a neutron star. If this is the  case, we find that the birth of a P-star could solve the supernova explosion problem leading 
to ordinary and superluminous core-collapsed supernovae.
Section~\ref{S3} is devoted to a  critical comparison of coalescing stellar-mass P-stars to the gravitational
event GW170817 and the observed electromagnetic follow-up. 
The gravitational waves from the inspiral and merger of two P-stars with masses exceeding  the neutron star limiting mass are qualitatively
discussed in Sect.~\ref{S4}. 
Finally, some concluding remarks are given in  Sect.~\ref{S5}.
\section{P-stars}
\label{S2}
\begin{figure}[t]
\centering
\includegraphics[width=0.80\textwidth,clip]{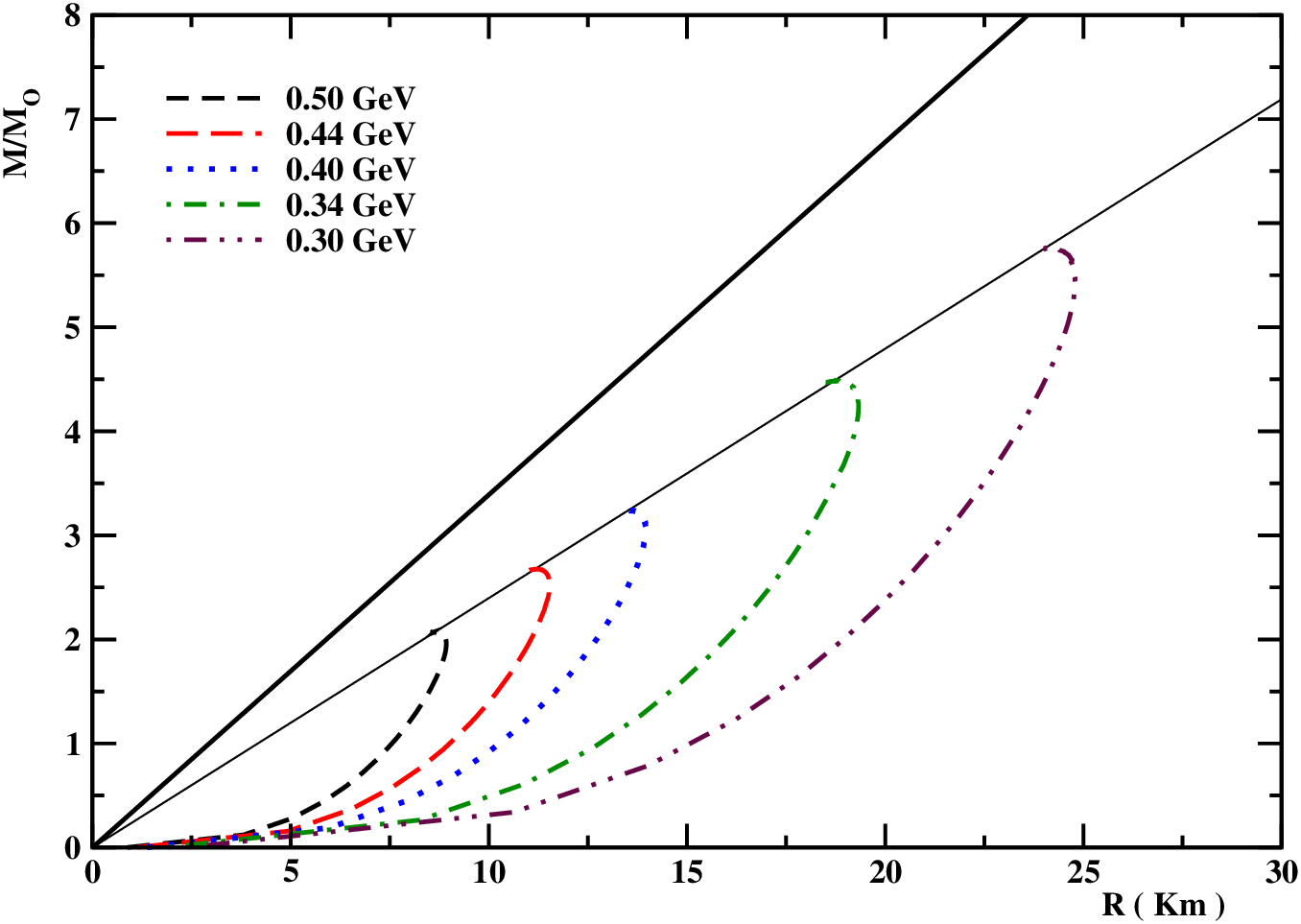}
\caption{\label{Fig1} Gravitational mass M, in units of the solar mass M$_{\odot}$, versus the stellar radius for P-stars with
different values of the chromomagnetic condensate field strength $\sqrt{gH}$ (in GeV). The tick solid line is the relation R=R$_S$,
where $R_S = 2 G M$ is the Schwarzschild radius; the thin solid line is the relation R = $\sqrt{2} \, R_S$.}
\end{figure}

Before going further, it can be useful to review the main arguments that led us to the proposal of P-stars as a viable alternative
to neutron stars and black holes. \\
P-stars are compact stars made of up and down quarks in $\beta$-equilibrium with electrons in an almost uniform chromomagnetic condensate.
We were led to investigate these  peculiar stellar objects from a remarkable property of the  vacuum of quantum chromodynamics.
Indeed, by means of  non-perturbative numerical simulations the QCD vacuum was probed by external constant chromomagnetic fields
directed along the third direction in the color space~\cite{Cea:2003,Cea:2005,Cea:2007}. It resulted that increasing the strength of the
applied external field led to a decrease of the deconfinement temperature $T_c$ such that it goes to zero at a certain critical field
 $gH_c$, where $g$ is the color gauge coupling and $H$ is the strength of the chromomagnetic field~\footnote{Note that through the
 paper we are adopting cgs units; however, in this Section we use the natural units of high-energy physics were
the Planck constant and  the speed of light  are set to one, $\hslash \; = \; c \; =  \; 1$. In these units the color coupling constant $g$ is
dimensionless and $\sqrt{gH}$ has dimension of energy. To switch to cgs units it suffices to replace $gH$ with $\hslash  c gH$. It is
also useful to point out that 1.0 GeV$^2 \, \simeq \,$  5.13 $\times$ 10$^{19}$ G.}.
 In other words, there is a critical field such that for $gH > gH_c$ the gauge system is in the deconfined phase.  This remarkable property
 of the QCD vacuum has been called the color Meissner effect since it is analogous to the Meissner-Ochsenfeld effect in ordinary
 superconductors (for a good account see, for instance, Ref.~\cite{Tinkham:1996}). On the other hand, it is well known that the
 QCD vacuum confines color charges at least at not too high temperatures.  We see, then, that  the color Meissner effect and the 
 existence of a critical chromomagnetic field must be compatible with the confining properties of the QCD vacuum. This led us  
 to suppose that the QCD vacuum behaves like a disordered chromomagnetic condensate. Indeed, at least in two spatial dimensions,
  long time ago Feynman~\cite{Feynman:1981} was able to show that the confining vacuum resembles a chromomagnetic condensate
disordered by the gauge invariance. Actually, the color confinement comes from both the existence of a mass  gap and the absence
of color long-range order. If the Feynman's picture extends to the case of three spatial dimensions, then the color Meissner effect is
compatible with the color confinement since a strong enough chromomagnetic field enforces long-range color order thereby
destroying the color confinement. \\
At low temperatures, that is the case of interest to us given that in relativistic compact stars the temperature is essentially zero everywhere,
it is widely believed that hadronic matter gets deconfined at densities well above the nuclear density. Within the above picture of the
QCD vacuum the most economic way to loose confinement is to restore the color long-range coherence by enforcing an almost uniform
chromomagnetic condensate. We were led, thus, to figure the deconfined hadronic matter as almost massless up and down quarks
immersed  in an uniform chromomagnetic condensate, named P-matter~\cite{Cea:2004a}. Taking into account that the deconfined quark
matter comes from nuclear matter, i.e. neutrons and protons, squeezed to densities above the nuclear density, we are led to quarks in chemical
equilibrium with electrons to ensure the electric charge neutrality. Obviously, the chemical equilibrium is assured by quark weak $\beta$-decays.
\\
Neglecting the electron and quark masses, the relevant thermodynamic potential is given by~\cite{Cea:2004a}:
\begin{equation}
\label{2.1}
 \frac{\Omega}{V} \; = \; - \, P \; = \; - \, \frac{gH}{4 \pi^2} ( \mu^2_u \; + \; \mu^2_d ) \; - \; \frac{\mu^4_e}{12 \pi^2} 
 \; + \; \frac{11}{32 \pi^2} \, (gH)^2  \; ,
\end{equation}
where $P$ is the pressure and $\mu_f$ ($f = u, d, e$) is the fermion chemical potential. The last term on the right hand size of Eq.~(\ref{2.1})
is the contribution to the thermodynamic potential of the chromomagnetic condensate. It is worthwhile to comment
on the origin of this term. Indeed, an almost uniform chromomagnetic condensate would contribute to the thermodynamic potential
with the vacuum energy. Usually, the vacuum energy is composed by the classical energy plus the quantum corrections. On the contrary,
in Eq.~(\ref{2.1}) there is only the contribution to the vacuum energy due to quantum fluctuations. This is due to the peculiar condensation
of tachyonic modes that completely cancel the classical term in the vacuum energy.  A full quantum-mechanical calculation within a gauge-invariant
variational approximation has been presented in Refs.~\cite{Cea:1987,Cea:1988} for the SU(2) gauge group. The term displayed in
Eq.~(\ref{2.1}) is our estimate for SU(3) that is the gauge group relevant to quantum chromodynamics. 
\\
Using  Eq.~(\ref{2.1}) one easily obtain the energy density~\footnote{Note that, since we set $c=1$, the energy density coincides with the 
mass density $\rho$.}:
\begin{equation}
\label{2.2}
\varepsilon   \; = \;   \frac{gH}{4 \pi^2} ( \mu^2_u \; + \; \mu^2_d ) \; + \; \frac{\mu^4_e}{4 \pi^2} 
 \; + \; \frac{11}{32 \pi^2} \, (gH)^2  \; .
\end{equation}
The chemical potentials are further constrained by the $\beta$-equilibrium condition:
\begin{equation}
\label{2.3}
 \mu_e \; + \; \mu_d  \; =  \; \mu_u  
\end{equation}
and charge neutrality:
\begin{equation}
\label{2.4}
 \frac{2}{3} \, n_u \; - \; \frac{1}{3} \, n_d   \; =  \;  n_e  \; ,
\end{equation}
where
\begin{equation}
\label{2.5}
 n_u  \;  \simeq \; \frac{1}{2 \pi^2} \, gH \, \mu_u   \; \; \; , \; \; \;  n_d  \;  \simeq \; \frac{1}{2 \pi^2} \, gH \, \mu_d 
\end{equation}
and 
\begin{equation}
\label{2.6}
 n_e  \;  \simeq \; \frac{1}{3 \pi^2} \,  \mu_e^3   \; \;  .
\end{equation}
Combining Eqs.~(\ref{2.3}) -  (\ref{2.6})  one gets the quark chemical potentials in terms of the electron chemical potential. After that, it is
easy to obtain the equation of state:
\begin{equation}
\label{2.7}
P   \; \simeq  \;  \varepsilon \; - \;   \frac{\mu^4_e}{6 \pi^2} 
 \; - \; \frac{11}{16 \pi^2} \, (gH)^2  \; .
\end{equation}
From the equation of state we find  the adiabatic sound speed:
\begin{equation}
\label{2.8}
 c^2_s \, = \, \frac{d P}{d \varepsilon } \, \simeq \,1 \, -  \;
\frac{1}{\frac{39}{2} \, + \, 18 \, \overline{\mu}^2_e \, + \,
\frac{15}{4 \, \overline{\mu}^2_e}} \; ,
\end{equation}
where $\overline{\mu}_e = \frac{\mu_e}{\sqrt{gH}}$  is the dimensionless electron chemical potential. It is interesting to observe that, irrespective
to the actual value of $\overline{\mu}_e$, the adiabatic sound  velocity is quite close  to the causal limit $c_s \, \simeq \, 1$. 
A large sound speed points to a dense matter with a stiff equation of state. Actually, reference to stiff or soft with regard to an equation of state
is a relative term. In fact, by stiffer it is meant that at given energy density the pressure is higher with respect to a softer equation
of state. Therefore, the stiffest equation of state that is compatible with causality is the one for which the speed of sound is the light speed.
This should imply that the equation of state   Eq.~(\ref{2.7})  is very stiff.  In general, one may assume that the pressure and the energy density
are connected by a power law of the form:
\begin{equation}
\label{2.9}
P \;  \sim  \;   \varepsilon^{\Gamma} 
\end{equation}
where
\begin{equation}
\label{2.10}
\Gamma \; = \; 1 \; + \; \frac{1}{n} 
\end{equation}
with $n$ a constant, which need not to be an integer, called the polytropic index~\cite{Tooper:1964}. Comparing   Eq.~(\ref{2.7})  
with  Eq.~(\ref{2.9}), we see that the equation of state of P-matter is very soft since $\Gamma \simeq 1$, while a stiff equation of state
requires $\Gamma > 1$. \\
Let us consider, now,  a static, spherically symmetric, relativistic star made of P-matter. To derive the differential equations of the star
we need to solve the Einstein field equations~\footnote{We shall follow the conventions adopted in Ref.~\cite{Glendenning:2000}.}:
\begin{equation}
\label{2.11}
R_{\mu\nu} \; - \;  \frac{1}{2} \, g_{\mu\nu} \, R  \; = \; - \, 8 \pi \, G \, T_{\mu\nu} 
\end{equation}
where   $R_{\mu\nu}$ is the Ricci tensor. The energy-momentum tensor  $T_{\mu\nu}$ can be written as:
\begin{equation}
\label{2.12}
 T_{\mu\nu}  \;  =  \; - P \, g_{\mu\nu} \; + \; (P \, + \, \varepsilon) u^{\mu} u^{\nu} \; , \;    u^{\mu} u_{\mu} \, = \, 1
\end{equation}
where  $u^{\mu}$ is the local fluid four-velocity. We are interested in the case of a static, spherically symmetric relativistic star. In a spherical
coordinate system at rest with respect to the star the metric can be written in the standard form:
\begin{equation}
\label{2.13}
 ds^2  \;  =  \;  \exp{(2 \nu(r))} \, dt^2  \; - \;  \exp{(2 \lambda(r))} \, dr^2  \; -  \; r^2 \, (d\theta^2 \; + \; \sin^2{\theta} \, d\phi^2 )  \; .
\end{equation}
Since we are assuming that the star is static, the non-zero components  of the energy-momentum tensor are:
\begin{equation}
\label{2.14}
 T_{0}^{0}   \;  =  \;  \varepsilon \; \; , \; \;   T_{i}^{i}   \;  =  \;  - \, P   \; \; , \; \;  i \, = \, 1,2,3 \; . 
\end{equation}
We need to find a solution of the Einstein field equations which is non-singular at the origin and that goes over to the Schwarzschild 
solution:
\begin{equation}
\label{2.15}
  \exp{(2 \nu(r))}   \;  =  \;  \exp{(- 2 \lambda(r))}   \; =   \; 1 \; - \; \frac{2 G M}{r} \; \; , \; \;  r\; \ge \; R 
\end{equation}
for a spherically symmetric distribution of total mass $M$ and coordinate radius $R$. Introducing:
\begin{equation}
\label{2.16}
 \exp{(- 2 \lambda(r))}   \; =   \; 1 \; - \; \frac{2 G }{r}  \, m(r) \; \; , 
\end{equation}
the Einstein field equations reduce to the Tolman-Oppenheimer-Volkov equations~\cite{Tolman:1939,Oppenheimer:1939}:
\begin{equation}
\label{2.17}
 \frac{d m}{d r} \, = \, 4 \; \pi \; r^2 \: \varepsilon(r) \;  \; ,
\end{equation}
\begin{equation}
\label{2.18}
 \frac{d P}{d r} \, = \, - \; \frac{G m(r) \varepsilon(r)}{r^2} \, \left [ 1 \, + \, \frac{4 \pi r^3 P(r)}{m(r)} \right ] \;
\frac{1 \, + \,\frac{P(r)}{\varepsilon(r)} }{1 \, - \, \frac{2 Gm(r)}{r}}  \;  \; .
\end{equation}
From  Eq.~(\ref{2.17}) we see that $m(r)$ can be interpreted as the total mass within a sphere of radius $r$. As concern the function $\nu(r)$
we obtain the following differential equation:
\begin{equation}
\label{2.19}
 \frac{d \nu(r)}{d r} \, = \, - \; \frac{G}{r^2} \;  \frac{ m(r) \, + \, 4 \pi r^3 P(r)}{1 \, - \, \frac{2 G m(r)} {r} }
\end{equation}
whose solution must match the exterior solution   Eq.~(\ref{2.15}). \\
To solve  the Tolman-Oppenheimer-Volkov equations we need  the equation of state $P=P(\varepsilon)$ as an input. In fact, one
imposes that at the origin $m(r=0)=0$, $\varepsilon(r=0)=\varepsilon_c$ and $P(r=0)=P(\varepsilon_c)$. After that, one can integrate
  Eqs.~(\ref{2.17}) and   Eq.~(\ref{2.18}) obtaining $m(r)$, $\varepsilon(r)$ and $P(r)$. The integration must end when $P(r=R)=0$,
  i.e. at the surface of the star the pressure vanishes. Then, the total mass of the star is $M=m(R)$. Finally, one can easily solve
  Eq.~(\ref{2.19}) with the boundary condition $\nu(r=R) = \frac{1}{2} \ln (1 - \frac{2 G M}{R})$ to ensure that the metric function
$\nu(r)$ will match continuously the  Schwarzschild metric outside the star. \\ 
We have solved numerically the Tolman-Oppenheimer-Volkov equations for different values of the chromomagnetic condensate $gH$.
The results  are displayed in Fig.~\ref{Fig1}. We see that, for a given value of the chromomagnetic condensate, the mass $M$ and radius $R$
of the P-star can be thought of as function of the central density $\varepsilon_c$. From  Fig.~\ref{Fig1} it is evident that the stellar mass $M$
increases with $\varepsilon_c$ until it reaches a maximum value at  $\varepsilon_c=\varepsilon_c^{max}$ where:
\begin{equation}
\label{2.20}
 \frac{d M(\varepsilon_c^{max})}{d \varepsilon_c} \;  =  \; 0 \; \; .
\end{equation}
 A further increase of $\varepsilon_c$ leads in a region where   $\frac{d M}{d \varepsilon_c} < 0$, so that the stellar system becomes unstable.
 Moreover, we see that, unlike neutron stars, P-stars do not admit a lower limit for the radius since for small masses one finds
 $R \sim M^{\frac{1}{3}}$. \\
Interestingly enough,  we find that:
\begin{equation}
\label{2.21}
\varepsilon_c^{max}  \;  \simeq   \;  0.22  \;  (gH)^2 \;  \; .
\end{equation}
 Therefore, on dimensional ground, from the Tolman-Oppenheimer-Volkov equations we get~\cite{Shapiro:1983}:
\begin{equation}
\label{2.22}
 M \; = \; \frac{1}{G^{3/2} gH} \; \;
 f(\overline{\varepsilon}_c) \;
\; \; \; ,  \; \; \;  R \; = \; \frac{1}{G^{1/2} gH} \; \;
g(\overline{\varepsilon}_c) 
\end{equation}
where $\overline{\varepsilon}_c \, = \, \varepsilon_c/(gH)^2 $. As a consequence we easily obtain~\footnote{Equations (\ref{2.22}) 
and (\ref{2.23}) in cgs units can be found in Ref.~\cite{Cea:2008}. }:
\begin{equation}
\label{2.23}
\frac{2 \; G \; M}{R} \; = \; 2 \;
\frac{f(\overline{\varepsilon}_c)}{g(\overline{\varepsilon}_c)} \;
\; \equiv \; h(\overline{\varepsilon}_c).
\end{equation}
Equations  (\ref{2.22}) and (\ref{2.23}) show  that by decreasing the strength of the chromomagnetic condensate  the mass $M$
 and radius $R$ of P-stars may increase without upper bounds, but the ratio $\frac{2  G  M}{R} $ depends only on  $\overline{\varepsilon}_c$.
Note that, for  $h(\overline{\varepsilon}_c) = 1$,  $R$ in Eq.~(\ref{2.23}) reduces to the Schwarzschild radius: 
\begin{equation}
\label{2.24}
R_S \; = \;  2 \, G \, M \; \; .
\end{equation}
On the other hand, it results that  $h(\overline{\varepsilon}_c)  < 1$, more precisely we find that  
$h(\overline{\varepsilon}_c)  \lesssim \frac{1}{\sqrt{2}}$. As a consequence, for the most compact P-stars we have (see Fig.~\ref{Fig1}):
\begin{equation}
\label{2.25}
R_{max} \; \simeq  \;  2 \, \sqrt{2} \,  G \, M \; \; .
\end{equation}
We are led to the remarkable consequence that our peculiar equation of state of degenerate up and down quarks in an almost
uniform chromomagnetic condensate allows the existence of finite equilibrium states for stars of arbitrary masses. In other words,
P-stars do not admit the existence of an upper limit to the stellar mass. \\
Our previous discussion on the equation of state makes evident that the chromomagnetic condensate  acts like a dynamical effective bag constant.
On the other hand, in general the chromomagnetic condensate strength in P-stars is bounded from above by the natural 
upper limit:
\begin{equation}
\label{2.26}
\sqrt{gH} \;  \lesssim \;   \sqrt{gH_c} \;  \; .
\end{equation}
From extensive non-perturbative numerical simulations of QCD on the lattice we infer~\footnote{ The range of values displayed in Eq.~(\ref{2.27})
 is our estimate of the lattice numerical results presented in Refs.~\cite{Cea:2003,Cea:2005,Cea:2007} extrapolated to the physical point
 where the mass of the pion is about 140 MeV.}:
\begin{equation}
\label{2.27}
 \sqrt{gH_c} \;   \simeq  \; 1.1  \; - \; 3.5 \; \;  {\text{GeV}} \; \; .
\end{equation}
Nevertheless, from Fig.~\ref{Fig1} we see that solar-mass P-stars require chromomagnetic condensate with strengths smaller
by about an order of magnitude with respect to  $\sqrt{gH_c} $. Actually, a more stringent bound can be obtained if we require
that P-stars are gravitationally self-bound, namely with positive  binding energy. To see this, we need the baryon number:
\begin{equation}
\label{2.28}
A =  4 \pi \, \int_0^R \frac{n_b(r)}{\sqrt{ 1 \, - \frac{2Gm(r)}{r}}} \, r^2 \, dr 
\end{equation}
where $n_b(r)=\frac{1}{3}(n_u(r)+n_d(r))$.  After that, the binding energy per nucleon is given by:
\begin{equation}
\label{2.29}
B \;  = \;  m_N \;  - \;  \frac{M}{A} 
\end{equation}
where $m_N \simeq 0.938$ GeV is  the nucleon mass. For a gravitationally self-bound P-star we require that $B > 0$. This leads to the
more stringent constraint:
\begin{equation}
\label{2.30}
\sqrt{gH} \;  \lesssim \;   0.45 \;  \text{GeV}  \; .
\end{equation}
It is interesting to note that in our proposal stable branch of equilibrium configurations can appear providing solutions of 
the Tolman-Oppenheimer-Volkov equations with the same mass but different radii giving rise to the so-called twin-star configurations.
For instance, for a P-star with canonical mass $M  \simeq 1.4 \; M_{\bigodot}$, we find:
\begin{equation}
\label{2.31}
\begin{split}
M  & \, \simeq \, 1.4 \, M_{\bigodot}  \; , \;  R \, \simeq \, 11.0 \, \text{Km} \; , \;  B  \simeq  15 \; \text{MeV} \;  \; , \; \sqrt{gH} \; \simeq  \;  
 0.40 \;  \text{GeV}  \, , 
 \\
M  & \, \simeq \, 1.4 \, M_{\bigodot}  \; , \;  R \, \simeq \, 14.0 \, \text{Km} \; , \;  B  \simeq  130 \; \text{MeV} \, , \, \sqrt{gH} \; \simeq  \; 
  0.34 \;  \text{GeV}  \, , 
\\
M  & \, \simeq \, 1.4 \, M_{\bigodot}  \; , \;  R \, \simeq \, 17.0 \, \text{Km} \; , \;  B  \simeq  210 \; \text{MeV} \, , \, \sqrt{gH} \; \simeq  \;  
 0.30 \;  \text{GeV}  \, .
\end{split}
\end{equation}
\begin{figure}
\vspace{-0.5cm}
\begin{center}
\includegraphics[width=0.80\textwidth,clip]{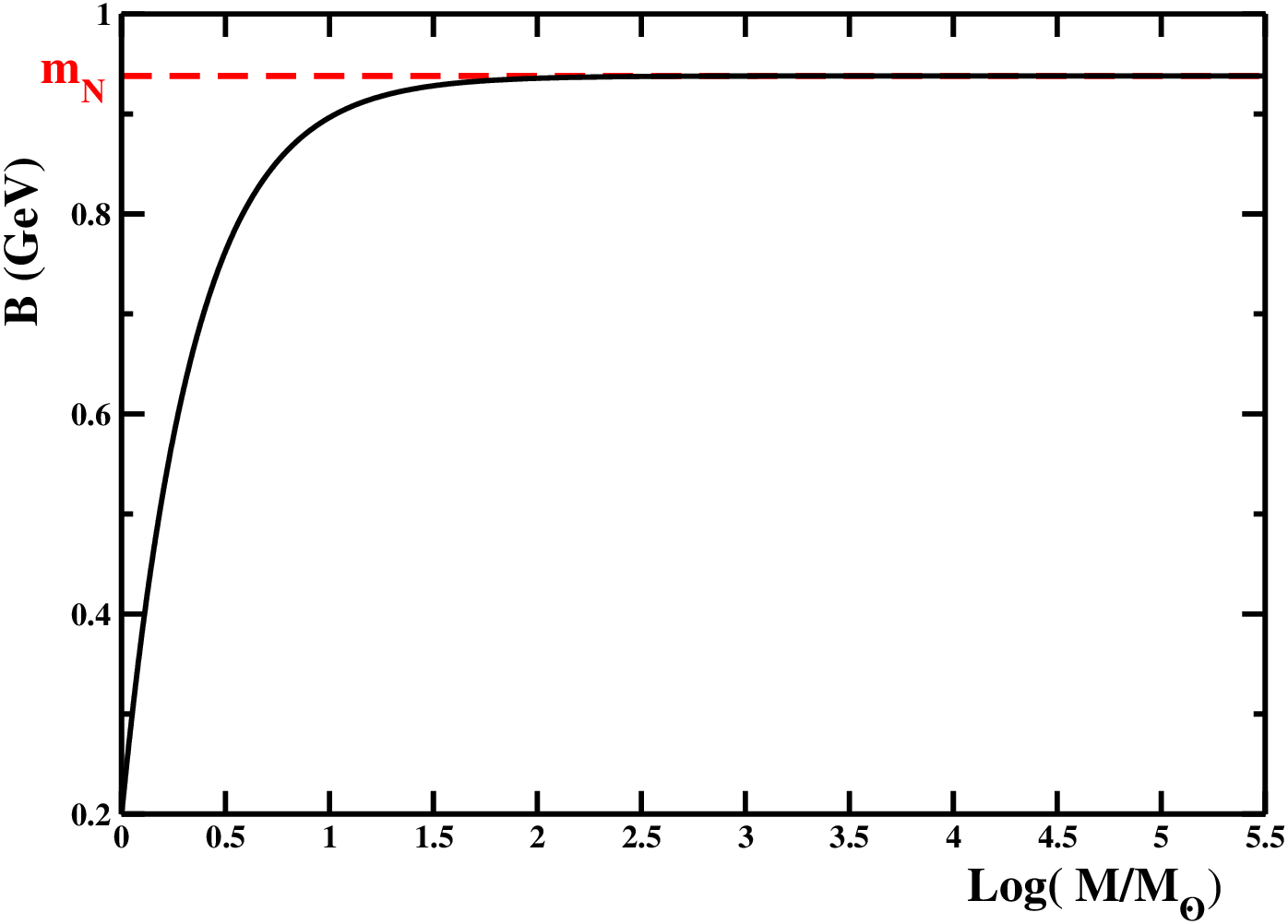}
\caption{\label{Fig2}  The binding energy (in GeV)  per nucleon for P-stars with limiting mass versus the decimal logarithm of the stellar
mass in units of the solar mass. The dashed line is the nucleon rest mass energy.}
\end{center}
\end{figure}
We, also, note that the binding energy tends to increase with the decrease of the strength of the chromomagnetic condensate. This can
be better appreciated looking at Fig.~\ref{Fig2} where we report the binding energy as a function of the stellar mass for the central energy
density corresponding to stable P-stars with the greatest allowed gravitational mass. Indeed, we see that the binding energy is a
monotonically increasing function of the mass $M$ that saturates at the nucleon rest-mass energy $m_N$ for sufficiently high masses
$M \gtrsim 10 \, M_{\bigodot}$.  Therefore, massive P-stars are the hardest objects in the Universe and they are practically not
deformable. Moreover, to a good approximation, the radius of a massive P-star is given by Eq.~(\ref{2.25}), so that for the compactness
we get:
\begin{equation}
\label{2.32}
\frac{G \; M}{R} \; \simeq \; \frac{1}{2 \, \sqrt{2}} \; \simeq \; 0.354 \; \; ,
\end{equation}
that is only slightly smaller than the maximum possible value consistent with causality. \\

Compact relativistic astrophysical objects are generally thought to be neutron stars or black holes. Nevertheless, we feel that there are
several observations that could point to a drastic revision of this standard paradigm. We would like to stress, once more, that this
matter can be settled down only by extensive precise astrophysical observations. In  fact, in our previous papers we performed
several studies aimed to ensure that P-stars were not in blatant contradiction with the huge available pulsar  phenomenology.
In the remainder of the present Section we take care of the dramatic recent observational progresses, due to the detection of gravitational waves 
from coalescing compact objects together with the observations of massive pulsars, to compare our P-stars with neutron stars. \\
The main theoretical problems that one faces dealing with neutron stars are due to the lack of a precise enough knowledge of the 
equation of state. Indeed,  a long-standing problem in nuclear physics is a reliable determination of the correct equation of state
for cold nuclear matter above the nuclear density (see, for instance, the reviews 
Refs.~\cite{Lattimer:2012,Baldo:2016,Lattimer:2016,Oertel:2017,Baym:2018}  and references therein). Fortunately, tight astrophysical
constraints on the equation of state come from the recent observations of massive pulsars and the detection of gravitational waves from
the merger of solar-mass binary compact stars. Firstly, from the observation of the massive pulsars
PSR J1614-2230 with mass $M  =  1.97 \, \pm \, 0.04 \,  M_{\bigodot}$~\cite{Demorest:2010}, 
PSR J0348+0432  with mass $M  =  2.01 \, \pm \, 0.04 \,  M_{\bigodot}$~\cite{Antoniadis:2013},
PSR J1600-3053  with mass $M  =  2.5^{+0.9}_{-0.7} \,  M_{\bigodot}$~\cite{Arzoumanian:2018},
PSR J2215+5135 with mass $M  =  2.27^{+0.17}_{-0.15} \,  M_{\bigodot}$~\cite{Linares:2018},
PSR J0740+6620 with mass $M  =  2.14^{+0.10}_{-0.09} \,  M_{\bigodot}$~\cite{Cromartie:2020},
it follows that one must require a rather stiff neutron-star equation of state. Moreover, the existence of massive pulsars imposes
severe constraints on the phases of dense hadronic matter. At the same time, a robust constraint on the equation of state emerges
from the detection of gravitational waves originating from the merger events GW170817~\cite{LIGO:2018} and
GW190814~\cite{LIGO:2020}. \\
\begin{figure}
\vspace{-1.3 cm}
\begin{center}
\includegraphics[width=0.80\textwidth,clip]{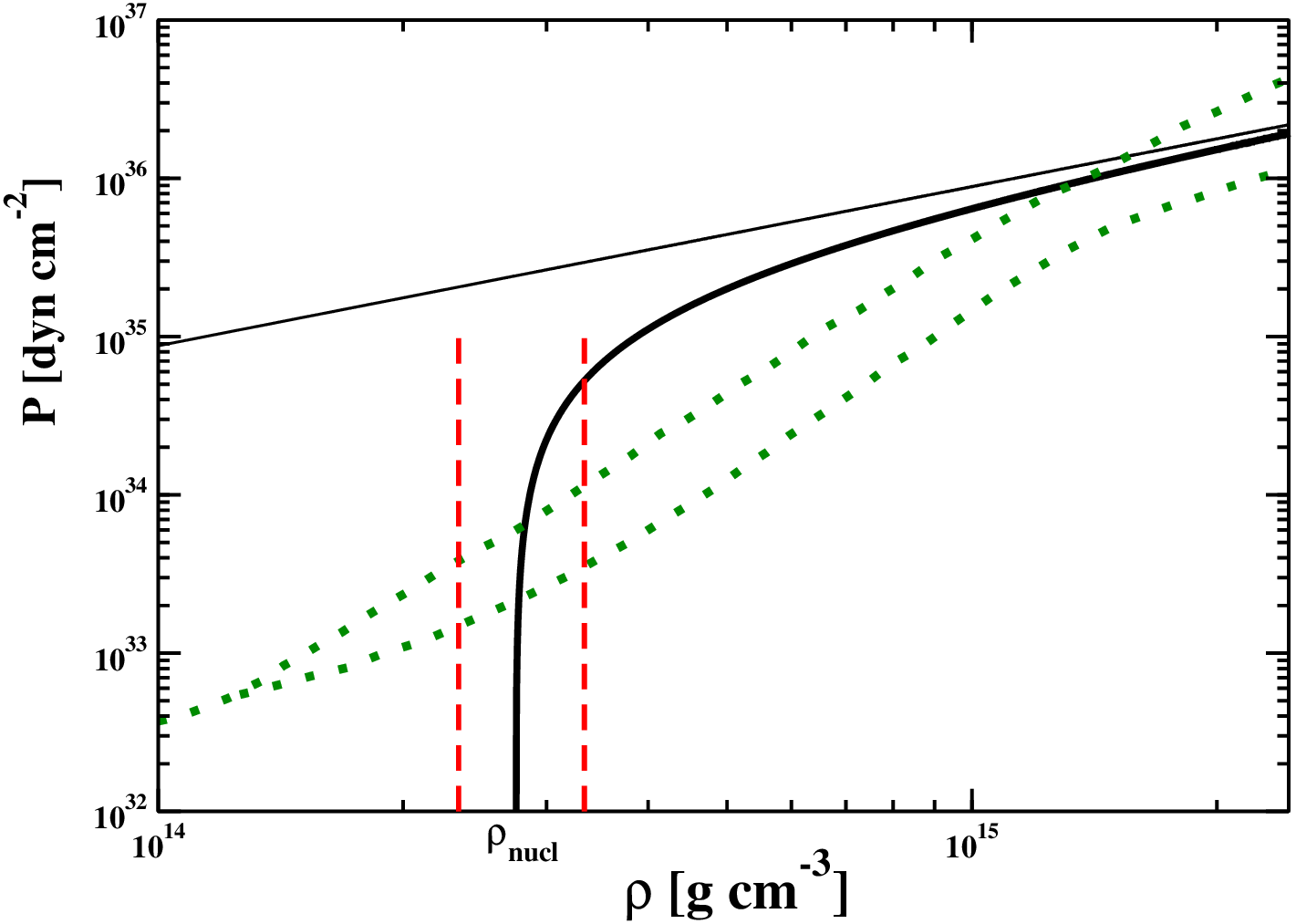}
\caption{\label{Fig3} 
P-star equation of state for $\sqrt{gH} = 0.40$ GeV (thin solid line) and $\sqrt{gH}=2.03$ GeV (thick solid line).
The dotted lines correspond to the 90 \%  credible level for the pressure as a function of the rest-mass density assuming that
the gravitational event GW170817 is due to the coalescing of two neutron stars~\cite{LIGO:2018}. The data have been taken from Fig.~8
of Ref.~\cite{LIGO:2020}. The vertical dashed lines represent the saturation density of the nuclear matter according to Eq.~(\ref{2.36}). }
\end{center}
\end{figure}
It is useful, now, to compare the P-star equation of state with the neutron-star equation of state allowed by the above robust astrophysical
constraints. To this end, we need to solve Eqs.~(\ref{2.1}) - (\ref{2.6}). From  Eqs.~(\ref{2.1}) - (\ref{2.6}) we readily get: 
\begin{equation}
\label{2.33}
 P \; \simeq \;  \varepsilon \;  - \; \frac{\mu^4_e}{6 \pi^2} 
 \; - \; \frac{11}{16 \pi^2} \, (gH)^2  \; ,
\end{equation}
\begin{equation}
\label{2.34}
 \varepsilon \; \simeq \;  + \; \frac{13}{4 \pi^2} \, \mu^4_e \; +  \; \frac{2}{\pi^2} \, \frac{\mu^6_e}{gH} \; 
 + \frac{5}{4 \pi^2} \, gH \,  \mu^2_e\; ,
\end{equation}
\begin{equation}
\label{2.35}
\mu_u \; \simeq \; \mu_e \; + \; 2 \,  \frac{\mu^3_e}{gH}  \; \; \; , \; \; \;  \mu_d \; \simeq \; 2 \, \mu_e \; + \; 2 \,  \frac{\mu^3_e}{gH}
\; \; .
\end{equation}
For a given value of the chromomagnetic condensate we may solve numerically Eq.~(\ref{2.34})  to get
$\mu_e=\mu_e(\varepsilon)$. After that, inserting  $\mu_e(\varepsilon)$ into   Eq.~(\ref{2.33}) we get $P=P(\varepsilon)$.  
In Fig.~\ref{Fig3} we display the equation of state of P-stars for densities of the nuclear matter relevant to stellar-mass 
compact stars. We checked that in the displayed density interval the P-star equation of state shows a very weak dependence on
the chromomagnetic condensate strength as long as $\sqrt{gH} \lesssim 0.45$ GeV.
We may, now, compare our equation of state to the tight constrained neutron-star equation of state reported in Ref.~\cite{LIGO:2018}
from the gravitational wave event GW170817 (see Fig.~\ref{Fig3}). 
In fact, assuming that GW170817 was due to the coalescence of a binary neutron star
system, in Ref.~\cite{LIGO:2018} it was obtained the 90 \% and 50 \% credible contours of the posterior in the pressure-density plane.
The constraints were calculated by assuming a spectral decomposition of the equation of state. The authors of Ref.~\cite{LIGO:2018}
employed the spectral parametrisation that expresses the logarithm of the adiabatic index $\Gamma$ (see Eq.~(\ref{2.9})) of the equation
of state as a polynomial of the pressure. The adiabatic index is then used to compute the energy density and the rest-mass density which
are inverted to give the equation of state. Then, the parametrized equation of state is stitched to the SLy equation of state~\cite{Douchin:2001}
below about half the nuclear saturation density~\cite{Haensel:1981} (see, also, Refs.~\cite{Baldo:2016,Lattimer:2016,Oertel:2017})
~\footnote{Note that 1 GeV$^4 = 2.086 \times 10^{35}$ dyn/cm$^2$ and 1 GeV$^4$/c$^2$ = 2.32 $\times 10^{14}$ gr/cm$^3$.}:
\begin{equation}
\label{2.36}
\rho_{nucl}  \; = \; 2.84 \; \pm \; 0.50 \; \times \; 10^{14}  \; \;  \frac{\text{gr}}{\text{cm}^3}   \; \;  \; .
\end{equation}
Moreover, it was imposed  the requirement that the equation of state must support neutron stars up to at least 1.97  $M_{\bigodot}$
as a conservative estimate based on the observations of massive pulsars. We have checked that similar results hold if one imposes 
the constraints arising from the gravitational wave event GW190814. \\
Looking at  Fig.~\ref{Fig3} we see that the P-star equation of state supports a pressure considerably higher than the neutron-star
equation of state at least in the interval of interest for the mass density. Under this aspect we can said that the P-star equation of state
is very stiff. On the other hand, the increase of the pressure with the density is very slow, in fact consistent with an adiabatic index
close to $\Gamma \simeq 1$, leading to the softest  equation of state.
\begin{figure}
\vspace{-0.5cm}
\begin{center}
\includegraphics[width=0.80\textwidth,clip]{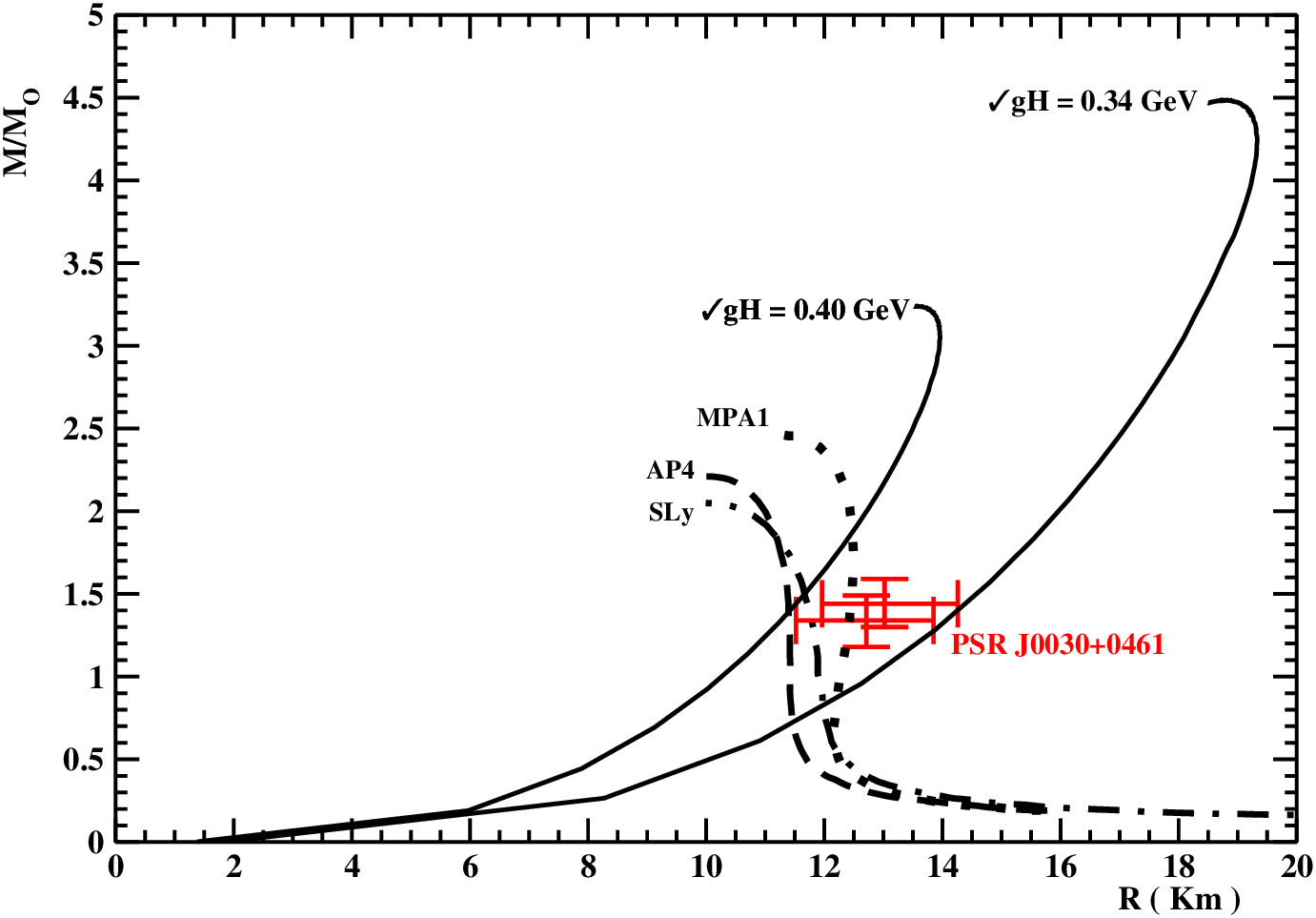}
\caption{\label{Fig4}  $M-R$ curves for P-stars with two different values of the chromomagnetic condensate and for
neutron stars corresponding to Sly, AP4 and MPA1 equations of state. For the SLy equation of state the data have been taken from Fig.~6
of Ref.~\cite{Douchin:2001}. For the equations of state AP4 and MPA1, see Fig. ~3 of Ref.~\cite{Lattimer:2012}. For the nomenclature
of the neutron-star equations of state we followed Ref.~\cite{Lattimer:2001}. The data points correspond to the precise mass and radius determination of the millisecond pulsar PSR J0030+0451 reported in Refs.~\cite{Riley:2019,Miller:2019}.}
\end{center}
\end{figure}
It is interesting to compare the equilibrium stellar configurations obtained by solving the Tolman-Oppenheimer-Volkov equations with the
P-star equation of state and with  selected representative nuclear equations of state stiff enough to allow massive 
neutron stars (see Fig.~\ref{Fig4}).
For concreteness, we report in Fig.~\ref{Fig4} the recent tight constraints on mass and radius of the millisecond pulsar PSR J0030+0451
obtained in Refs.~\cite{Riley:2019,Miller:2019} by using  the Neutron Star Interior Composition Explorer (NICER) data~\cite{Raaijmakers:2019}.
Interestingly enough,  the observed mass and radius values of the nearby rotational-powered millisecond pulsar J0030+0451 are consistent
with both P-stars and neutron stars. Moreover,  the total binding energy, that contains both the gravitational and nuclear binding energies, for
neutron stars corresponding to the SLy, AP4 and MPA1 equations of state are $\lesssim 100$ MeV (see Table 1 in Ref.~\cite{Jong:2019}).
Considering that the binding energies for P-stars and neutron stars are comparable, we see that, even the precise experimental
determinations of masses and radii are nor able, at least up to now, to ascertain with certainty the true nature of solar-mass compact
objects. There is, thus, the intriguing possibility that P-stars could coexist with neutron stars. This leads us to address the problem
of the mechanisms responsible for the formation of P-stars. \\
There is an almost unanimous opinion that neutron stars are formed in core-collapsed
 supernovae~\cite{Woosley:1986,Bethe:1988,Arnett:1989,Bethe:1990,Arnett:1996,Janka:2012,Scholberg:2012,Kotaka:2012,Burrows:2013,Foglizzo:2015,Smartt:2015,Mirizzi:2016,Janka:2016,Muller:2016}. Actually,  it is widely believed that neutron stars 
 are formed when a massive star with $ M \gtrsim 8 \,  M_{\bigodot}$
runs out of fuel and collapses. The central region of the star, the core, collapses crushing  together every proton and electron  into  a neutron,
When matter density exceeds about $2.0 \times 10^{14}$ gr/cm$^3$ all nuclei are dissolved into an  homogeneous system composed enterely
of neutrons since the admixture of protons must be accompanied by an equal number of electrons which contribute a large amount of kinetic
energy to the system. Such a nearly pure neutron system is called neutron matter. \\
When the neutron-matter density reaches about the nuclear saturation density $\rho_{nucl}$ (see Eq.~(\ref{2.36})), further compression is resisted 
by the repulsive hard-core of the nuclear interaction. The pressure, then, rises faster than the gravitational force so that the neutron-matter
core stops contracting leading to the core bounce. The inner region of the neutron matter that rebound consists of about $1.0 \, M_{\bigodot}$
occupying a region of linear size of order $10$ Km. The rebounding  inner core gives rise to a shock wave that continues moving out and
eventually exiting the collapsed core with enough energy, about $10^{51}$ erg, to eject the rest of the star with high velocity. This is basically
the so-called prompt explosion mechanism for ordinary supernovae characterised by the canonical explosion energy $E_{SN} \sim 10^{51}$ erg.
The gravitational energy gained from the collapse of the core seems large enough to account for the observed energy of the ejecta:
\begin{equation}
\label{2.37}
E_G  \; \sim  \; \frac{G \, M^2}{R} \; \sim 10^{53} \; \text{erg}  
\end{equation}
for $M \sim 1.0 \, M_{\bigodot}$ and $R \sim 10$ Km. The main difficulty of the  supernova theory is to explain how this gravitational energy
is transferred to the stellar envelope to reverse the infalling motion into an explosion. Indeed, the formation of the compact core of neutron
matter releases over hundred times more energy in neutrinos, due to the neutronization in which an electron is captured by a proton to produce
a neutron and a neutrino, than the typical energy of a supernova explosion. Moreover, instead of accelerating outward the shock experience
severe energy losses. So that the shock front stalls at about $10^2$ Km. It was soon clear, therefore, that the prompt bounce-shock mechanism
cannot succeed. Since the formation of the compact central core released about $10^{53}$ erg in neutrinos, it has been suggested that
neutrinos  could be the 
decisive  agents for powering the supernova explosion. However, even after considerable progresses in the understanding of the processes
that play a role during stellar collapse and explosion, the neutrino-driven mechanism is not yet finally established as the solution
of the supernova problem. We see, then, that the question why and how core-collapse supernovae explode is still one of the most important
problem of stellar astrophysics. We feel that the solution of this problem could be reached by following convincing alternatives based
on well-justified assumption concerning the relevant physics for the high-density nuclear matter. \\
We said that in core-collapsed supernovae the gravitational implosion of the inner core of the stellar matter leads to the formation of a 
neutron matter core, called the protoneutron star, where the density reaches about the nuclear saturation density. At this density the nucleons
come to a close contact such that the nuclear wavefunctions begin to overlap. It is conceivable, then, that the neutron matter turns relatively
smoothly into quark matter by a further increases of the density. It should be stressed, however, than the description of nuclear matter at high
densities is one of the main challenges of the strong interaction physics. In fact, the lack of a satisfying knowledge of the structure of
the confining QCD vacuum does not allow to characterise  from first principles the dense hadronic matter at supranuclear density. Nevertheless,
we may take advance from the informations gained after extensive non-perturbative numerical simulations in lattice QCD to attempt a plausible
description of the transition of the hadronic matter into quark matter at high enough densities. We have seen that the color Meissner effect 
led us to picture the QCD vacuum as a disordered chromomagnetic condensate. We may visualise the QCD vacuum as a collection of small
chromomagnetic domains with linear size $\sim 1/\sqrt{gH}$ and chromomagnetic field strength $\sim gH$ which fluctuate both in color
and spatial directions. The gauge invariance of the vacuum assures that one can rotate a given chromomagnetic domain in the spatial and
color directions without cost in energy. The confinement of quarks is assured from the lost of quantum coherence of the quark wavefunction
over a distance of order $\sim 1/\sqrt{gH}$. It is easy to convince oneself that the color Meissner effect is recovered if $gH \simeq gH_c$.
Now, if two nucleons are close enough that the nuclear wavefunctions have a sizeable overlap, then, in principle, the quark wavefunction can
spread over the two nucleons thereby reducing his kinetic energy. This means that the quarks must retain quantum coherence at least over
a distance of about  $\sim 2/\sqrt{gH_c}$. This, in turns, is assured if the chromomagnetic domains are ordered in the same color and spatial
directions at least for distance $\sim 2/\sqrt{gH_c}$. Note that the absence of an energy barrier for the orientation of the chromomagnetic domains
assures that this process is energetically favoured. Obviously, if a macroscopic number of nucleons is involved, then there is a smooth transition
of the nuclear matter at high enough density to a quark matter made of valence quarks (up and down) immersed in an almost uniform
chromomagnetic condensate with strength $gH_c$. We are led, then, to assume that in the inner core of the protoneutron star there is a transition
from the neutron matter to our peculiar quark matter (P-matter). However, the assumed smooth phase transition can be realised only if
the density and pressure vary continuously from the neutron-matter phase to the quark-matter phase. Remarkably, if we assume
\begin{equation}
\label{2.38}
\sqrt{g H_c}   \; \simeq  \;  2.03  \;  \text{GeV}  
\end{equation}
we find (see Fig.~\ref{Fig3}) that the P-star equation of state matches the  neutron-matter equation of state near the nuclear saturation
density $\rho_{nucl}$. It is worthwhile to point out that it is a non-trivial consistency check of our proposal that our estimate of the
critical chromomagnetic field strength, Eq.~(\ref{2.38}), is compatible with the lattice determination, Eq.~(\ref{2.27}). Looking at 
Fig.~\ref{Fig3}, we see that the transition into the quark matter stiffens the equation of state so that the pressure increases by an order
of magnitude with respect to the neutron-matter equation of state. This means that the inner quark core may sustain a further compression
leading to a very compact quark core. To see this, we have solved the Tolman-Oppenheimer-Volkov equations with the chromomagnetic field
strength fixed by Eq.~(\ref{2.38}) to get for the limiting mass:
\begin{equation}
\label{2.39}
R  \, \simeq \, 0.53 \, \text{Km} \; , \;   M \, \simeq \, 0.13 \, M_{\bigodot} \; , \; 
 A \, \simeq \, 3.48 \times 10^{55} \; , \;   B  \, \simeq  \, - 3.09 \,\text{GeV}  \;  .  
\end{equation}
From  Eq.~(\ref{2.39}) we infer the total gravitational energy:
\begin{equation}
\label{2.40}
E_G \; = \; - \, B \, A \; \simeq \;  1 .7  \times 10^{53} \; \text{erg}  \;  .  
\end{equation}
The  gravitational energy gained for the formation of the quark protostar can be completely converted into the kinetic energy of the shock wave
created by the inner core bounce. Indeed, the quark-matter  core arises from the neutron matter so that $n_d \simeq 2 \, n_u$ and there is no
sizeable energy loss due to weak-decay neutrino emissions. In the rapid expansion of the quark core the strength of the chromomagnetic
 condensate decreases according to $\sqrt{gH} \sim 1/R(t)$. At the same time, the neutron matter compressed by the shock wave is fused into
  quark matter. In this way both the mass and radius of the quark protostar increase until the star reaches a stable branch of equilibrium 
  configurations leading to the final stable P-star with  $M \simeq 1.0 \, M_{\bigodot}$ and $R \simeq 10$ Km. Note that the final stable
P-star is characterised by a chromomagnetic condensate reduced by an order of magnitude with respect to  $\sqrt{gH_c}$, in accordance
with  Eq.~(\ref{2.30}).  In the  adiabatic expansion of the quark core a fraction of the kinetic energy of the shock wave is dissipated into the
dissociation of nucleons in quarks. From the phenomenology of hadronization by coalescence or recombination of quarks we estimate that
the lost of kinetic energy is about 10 MeV per nucleon. Considering that a canonical P-star with  $M \simeq 1.0 \, M_{\bigodot}$,
 $R \simeq 10$ Km has a baryon number $A \sim 10^{57}$, we have:
\begin{equation}
\label{2.41}
E_{diss}  \; \sim \;  10^{52} \; \text{erg}  \; \ll \; E_G  .  
\end{equation}
It is clear, thus, that the prompt bounce-shock mechanism, where the hydrodynamic shock front from the quark core bounce directly,
is able to initiate the supernova explosion leading to successful supernovae with total energy up to $10^{53}$ erg.
It is remarkable that our proposal not only solves the supernova problem, but it could also account for superluminous supernovae,
i.e. a rare subtype of supernovae that are much brighter than ordinary stellar explosions (for
recent reviews, see Refs.~\cite{Howell:2017,Moriya:2018,Gal-Yam:2019}).
\section{GW170817}
\label{S3}
On 17 August 2017, the advanced LIGO and Virgo observed for the first time the merger of two solar-mass compact stars
GW170817~\cite{LIGO:2017d,LIGO:2017f,LIGO:2018}. The  total mass of the binary system is about  $2.73 \, M_{\bigodot}$,
the mass of the heavier component is around  $1.16 - 1.60 \, M_{\bigodot}$ with the lower-spin priors, while it can approach
 $1.89 \, M_{\bigodot}$ with high-spin priors~\cite{LIGO:2019}. Now, gravitational waves from such events are strongly influenced 
 by masses and tidal deformabilities, i.e. a parameter quantifying how the compact star deforms when an external gravitational
 field is applied. Since the deformability depends on the matter in the star, the gravitational wave observations provide an important
 means to study the equation of state of dense matter. In fact, assuming that the compact object in the binary system are neutron stars,
 we have already discussed in the previous Section the constraints on the neutron-star equation of state. Presently, we explore
 what further inferences we can make about the structure of neutron stars and P-stars from the gravitational wave event GW170817. \\
The early part of the gravitational wave signal  of binary-star inspirals yields robust information on the structure of the stars. Indeed,
the influence of a star's internal structure on the gravitational waveform is characterised by a single parameter $\lambda$, called
the tidal deformability, that measures the star quadrupole deformation in response to the perturbing tidal field produced by
the companion star. Therefore, one can write~\cite{Flanagan:2008,Hinderer:2008,Hinderer:2009}:
\begin{equation}
\label{3.1}
\mathcal{Q}_{ij}  \; = \; - \; \lambda \; \mathcal{E}_{ij}   \;  \; ,  
\end{equation}
where $\mathcal{Q}_{ij}$ is the quadrupole moment of the star and  $\mathcal{E}_{ij}$ the external tidal field generated by the companion
star~\cite{Thorne:1998}.  The constant $\lambda$ can be expressed in terms of the dimensionless Love number~\cite{Love:1909}.
Let $k_2$ be the quadrupole tidal Love number, then we can write\cite{Flanagan:2008,Hinderer:2008,Hinderer:2009}:
\begin{equation}
\label{3.2}
 \lambda \;  = \; \frac{2}{3} \; \frac{k_2 \, R^5}{G}  \;  \; ,  
\end{equation}
where, henceforth, we restore cgs physical units. The tidal Love number $k_2$ depends strongly on the star compactness:
\begin{equation}
\label{3.3}
\mathcal{C} \;  = \;   \frac{G \, M}{c^2 \, R}    
\end{equation}
where  $M$ and $R$ are the mass and radius of the star respectively. It can be seen that $k_2$ vanishes at the compactness
of a Schwarzschild black hole  $\mathcal{C}   =  \frac{1}{2}$. The tidal deformations are more usefully described through the
dimensionless deformability:
\begin{equation}
\label{3.4}
\Lambda \;  = \; \frac{c^{10} \, \lambda}{G^4 \, M^5} \; = \; \frac{2}{3} \, k_2 \, \left ( \frac{c^2  \, R}{G \, M } \right )^5  \;  \; ,  
\end{equation}
In general, the tidal deformabilities affect the gravitational wave signal of a coalescing binary system with a phase-shift that can be
efficiently extracted only in the late part of the inspiral. Moreover, it turns out that the tidal phase corrections to the  leading order
depend on the effective tidal deformability~\cite{Flanagan:2008}:
\begin{equation}
\label{3.5}
\tilde{\Lambda} \;  = \; \frac{16}{13} \;
 \frac{(M_1 + 12 M_2)M_1^4 \Lambda_1 + (M_2 + 12 M_1)M_2^4 \Lambda_2 }{(M_1 + M_2)^5} 
\end{equation}
where $M_{1,2}$ and $\Lambda_{1,2}$ are the masses and tidal deformabilities of the binary system. \\
In the gravitational wave event GW170817 there was not evidence of tidal deformability effects leading to constraints on $\Lambda_1$ and
$\Lambda_2$ that disfavour equations of state that predict less compact stars. More precisely, assuming an uniform prior on $\tilde{\Lambda}$,
the authors of Ref.~\cite{LIGO:2017d} placed a 90 \% upper limit 
$\tilde{\Lambda} \le 900$ in the low-spin case and    $\tilde{\Lambda} \le 700$ in the high-spin case. Since $\Lambda  \sim k_2 \, (R/M)^5$,
the upper limit on the tidal deformability can put a stringent limit on the radius of the binary star. To see how this works, at least
qualitatively, let us assume $M_1 \simeq M_2 = M$, $\Lambda_1 \simeq \Lambda_2 = \Lambda$. In the high-spin scenario, we get
$M \simeq  1.41 \, M_{\bigodot}$ and $\Lambda \lesssim 440$  at 50 \% of the
probability density (see Fig.~5, right panel in Ref.~\cite{LIGO:2017d}).  For neutron stars with realistic equations of state one has 
$k_2 \simeq 0.05 - 0.15$~\cite{Hinderer:2008,Hinderer:2009,Hinderer:2010,Postnikov:2010}.
Assuming $k_2 \simeq 0.10$, we obtain:
\begin{equation}
\label{3.6}
M \; \simeq \;  1.4 \, M_{\bigodot}  \; \; \; , \; \; \; R \; \lesssim \; 12 \; \text{Km} \; \; .
\end{equation}
\begin{figure}
\vspace{-0.5cm}
\begin{center}
\includegraphics[width=0.80\textwidth,clip]{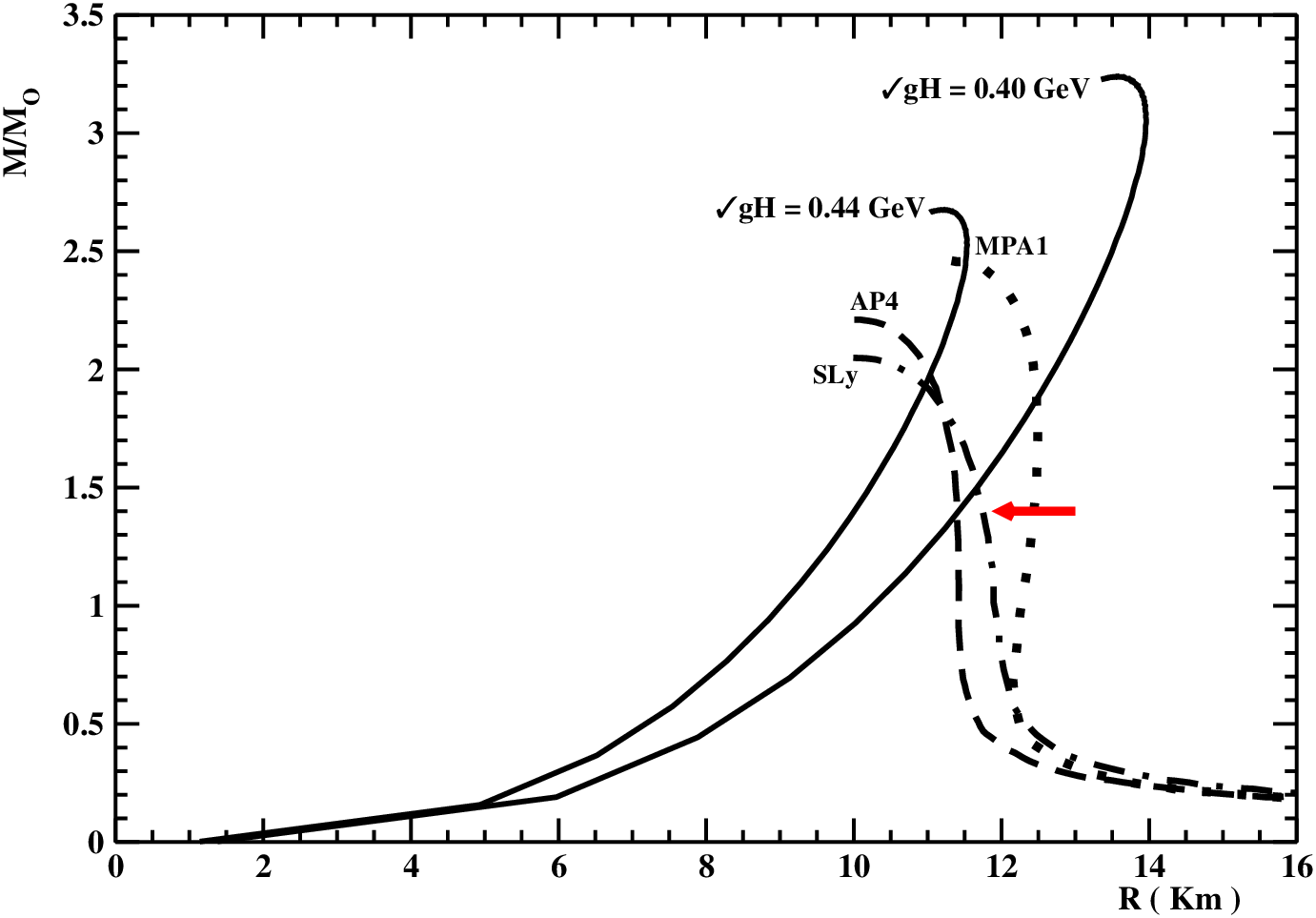}
\caption{\label{Fig5}  
Stellar sequences for neutron stars with realistic equations of state and P-stars with $\sqrt{gH} \simeq 0.40$ GeV and $0.44$ GeV.
The arrow indicates the upper limit Eq.~(\ref{3.7}).}
\end{center}
\end{figure}
Indeed, a more refined treatment gives~\cite{De:2018,Radice:2019}:
\begin{equation}
\label{3.7}
M \; \simeq \;  1.4 \, M_{\bigodot}  \; \; \; , \; \; \; R \; \lesssim \; 13 \; \text{Km} \; \; .
\end{equation}
In Fig.~\ref{Fig5} we compare the upper limit Eq.~(\ref{3.7}) to neutron-star equations of state  admitting limiting masses above 
$ 2 \,  M_{\bigodot}$. We also display the P-star equation of state for two representative values of the chromomagnetic condensate.
Even though the constraint on the radius is still consistent whit the neutron-star equations of state, it has been recently pointed out
that there is some hints of a tension between nuclear physics and astrophysical observations~\cite{Biswas:2020}.
In addition, there are also constraints on the radius obtained by means of $X$-ray spectral modelling of transiently accreting
and bursting pulsars~\footnote{For a detailed review that includes an illustration of $X$-ray modelling techniques and associated
uncertainties, see Ref.~\cite{Ozel:2016}.} that seems to favour  rather soft neutron-star equations of state. To illustrate better
this point, we briefly discuss few explanatory examples. 
\begin{itemize}
 \item
Since its discovery Her X-1, a binary $X$-ray pulsar, has been widely studied. Following Ref.~\cite{Wasserman:1983} a semi-empirical
mass-radius relation may be derived from several external properties of $X$-ray pulsars. In fact, the authors of Ref.~\cite{Reynolds:1997a}
argued  that the hypothesis that Hercules X-1 was a neutron star was not disproved provided that one adopts a very soft equation of
state. In particular, these authors, using the mass estimate $ M = 1.5 \pm 0.3  \,  M_{\bigodot}$ and the revised distance
$d = 6.6 \pm 0.4$ kpc~\cite{Reynolds:1997b}, found that $R \lesssim 10$ Km even assuming  Eddington luminosities.
\item 
From observations of the quiescent low-mass $X$-ray binaries X7 and X5 in the globular cluster 47 Tuc, the authors of Ref.~\cite{Bogdanov:2016}
inferred that these measurements strongly favoured radii in the $9.9 - 11.2$ Km range for a $\sim 1.5 \, M_{\bigodot}$ neutron star
and pointed to a dense matter equation of state somewhat softer than the nucleonic ones. 
\item
Using time-resolved spectroscopy of thermonuclear $X$-ray bursts observed from SAX J1748.9-2021, a transient $X$-ray binary located
in the globular cluster NGC 6440, the authors of Ref.~\cite{Guver:2013} inferred:
\begin{equation}
\label{3.8}
\;  \; R \; =  \; 8.18 \; \pm \; 1.62 \; \text{Km} \; \;  \; , \; \; \;  \;  M \; = \;  1.78  \; \pm \; 0.3 \; M_{\bigodot}  \; \; 
\end{equation}
or
\begin{equation}
\label{3.9}
\; \; \;  R \; =  \; 10.93 \; \pm \; 2.09 \; \text{Km} \; \; \;  , \; \; \; M \; = \;  1.33  \; \pm \; 0.33  \; M_{\bigodot} \; \; , 
\end{equation}
concluding that the pulsar radii were in the $ 8 - 11$ Km range.
\end{itemize}
The above rather incomplete discussion points to an intrinsic tension for the equations of state of dense matter that need to be
both stiff and soft~\cite{Drago:2019}. On the other hand, for P-stars there are no problems since, as shown in Fig.~\ref{Fig5},
our peculiar equation of state for quark matter allows canonical compact stars with  $R \simeq   10 $  Km,  $M \simeq 1.4  \, M_{\bigodot}$.
\\
We cannot conclude the present Section without mentioning that the most remarkable aspect of the gravitational event GW170817
has been the observation of the short Gamma-Ray Burst GRB170817A, delayed by about two seconds with respect to the determined 
merger time, and the numerous follow-up observations of electromagnetic radiation from what are thought to be the material
ejecta of the binary merger, the kilonova AT2017gfo~\cite{LIGO:2017f,LIGO:2017g} (a fair complete list of references accounting the huge
literature on this subject can be found in Ref.~\cite{Lazzati:2020}). There is an almost unanimous consensus that these observations
demonstrate without doubts that the binary compact stars involved in the gravitational event GW170817 are neutron stars.
In fact, when a neutron star collides with another neutron star, most of its material ends up within the remnant compact object, but
a small fraction of the neutron star may be dynamically ejected from the system and/or form an accretion disk around that compact
object. It is believed that these debris fuel the electromagnetic counterparts observed in the gravitational event GW170817.
Actually, as much of these processes are still poorly understood, here we must rely on phenomenological arguments widely used
in the literature. In particular, it is agreed that an enormous amount of energy is deposited in the polar region near the surface of 
the compact remnant which, then, launches a highly collimated, ultra-relativistic jet that results in a short Gamma-Ray Burst, as
confirmed by GRB170817A (see, eg, Refs.~\cite{Berger:2014,D'Avanzo:2015,Rosswog:2015,Dai:2017} and references therein).
However, it should be remarked that GRB170817A is several orders-of-magnitude fainter than ordinary short Gamma-Ray Bursts. This
peculiar  faintness has been interpreted as most likely due to the large off-axis viewing angle. In fact, through Very Long Baseline
Interferometry it resulted that the compact radio source associated with GW1708017 exhibits superluminal motion indicating the 
presence of an energetic and narrowly collimated jet with a rather small opening angle and observed from a viewing angle
of about 20 degrees~\cite{Mooley:2018}. 
In addition to the Gamma-Ray Burst and its afterglow, the detection of the transient AT2017gfo seems to give further support to the
binary neutron star scenario. Indeed, the kilonova AT2017gfo is thought to represent a robust electromagnetic counterpart to the
gravitational wave signal of binary neutron star merger. This is because the ejection of neutron rich material from binary
neutron star mergers is essentially inevitable. As a matter of fact, it has been suggested long time ago~\cite{Li:1998} that
kilonovae are thermal transients powered by the radioactive decays of heavy neutron-rich nuclei, which have been synthesised by
rapid neutron capture processes (r-process) in neutron-rich ejecta from neutron star mergers (for up to date reviews, see
Refs.~\cite{Tanaka:2016,Yu:2019,Metzger:2020}). To date, various models have been proposed to explain  the afterglow emission, but, so far,
there is no convincing global model. In the case of the kilonova AT2017gfo, there are indications that multi-component kilonova models
are necessary to interpret the observed light curves. Altough many details still have to be clarified, like the exact amount of the ejecta
or their composition, it turned out that a single-component kilonova is able to broadly reproduce the photometric light curves well.
The single-component fit is consistent with a large ejecta mass $M_{ej} \simeq 0.05  M_{\bigodot}$ and blue component (lanthadine-poor).
The velocity distribution is rather broad with $v_{ej} \simeq 0.06 - 0.30 \, c$. \\
The amount of the ejected mass is an important parameter since it depends on the neutron-star equation of state. In fact, for stiff equations
of state the ejected masses range from $10^{-3}  M_{\bigodot}$  to $10^{-2}  M_{\bigodot}$, while for soft equations of state
 $M_{ej} \gtrsim 10^{-2} M_{\bigodot}$~\cite{Bauswein:2013}. We see, thus, as for tidal deformability it emerges some tension with
 the neutron-star equation of state considering that the detection of massive pulsars required rather stiff equations of state. Nevertheless,
 for a precise comparison with observations, the complexity of the involved phenomena requires numerical simulations of binary compact
star mergers. Actually, despite decades of observations, the physics of the compact remnant powering the jet, the stability of the jet
and what is responsible for their collimation and acceleration, the amount and composition of the ejected mass remain  open
theoretical challenges. Moreover, considering that magnetic fields are widely recognised to play a fundamental role~\cite{Ciolfi:2020a,Ciolfi:2020b},
ultimately ab-initio magnetohydrodynamic simulations in full General Relativity are needed to constrain the models used to interpret
the electromagnetic counterparts of coalescing compact star gravitational events. \\
We have seen that the interpretation of the gravitational wave event GW170817 as coalescing neutron stars is qualitatively consistent with
theoretical expectations, but it gives rise to problematic tension with respect to realistic neutron-star equations of state severely 
constrained by recent astrophysical observations. On the contrary, if we assume that GW170817 is due to the coalescing
of a binary P-star system, then the emerging picture  looks more promising. Firstly,  we have already seen that there is no tension
between the P-star equation of state and the upper limit on the effective tidal deformability sets by the gravitational wave event GW170817.
On general grounds, we showed~\cite{Cea:2006} that there is a natural mechanism to endow P-stars with almost dipolar magnetic
field. Therefore, in the coalescence of two P-stars the resulting turbulent magnetic fields should play a fundamental role both in accelerating
and collimating relativistic flows. As in neutron stars, it is inevitable that in the violent merger of binary P-stars some stellar material
will be ejected. Now, for P-stars the stellar material is composed by up and down quarks with a small admixture of electrons. Since
the quarks have (fractional) electric charge, they can directly interact with the turbulent magnetic fields. It is conceivable, then, that
the quark matter can produce a highly collimated, ultra-relativistic jet near the polar region of the compact remnant that, in our case, is
a rapidly rotating P-star with mass $M \simeq 2.7 \, M_{\bigodot}$. One of the crucial properties of the strong force is the color
confinement, namely the fact that color charged quarks cannot exist as free particles outside the hadrons. As a consequence,
the ultra-relativistic quarks hadronize giving rise to a  narrow jet of hadrons. Indeed, these processes are routinely observed
in high-energy hadron collisions. In addition, in the collisions of the P-stars a small fraction of the stellar material may be dynamically ejected.
The amount  of ejected matter depends on the equation of state. We already observed the P-stars are characterised by a very soft equation
of state (adiabatic index $\Gamma \simeq 1$), so that we expect for the ejected matter $M_{ej} \lesssim 10^{-1} M_{\bigodot}$.
Obviously, the unbound matter is composed by up and down quarks with a small fraction of electrons  with, probably,  sub-relativistic
velocities. Once the quarks are outside the stars, by color confinement, they must hadronize.  In a densely populated system, quarks
could directly recombine into hadrons. This coalescence mechanism intuitively describes hadronization assuming that quarks
join into a hadron when they are close to each other in space and travelling with similar velocities (see, for instance, Ref.~\cite{Fries:2008}
and references therein). Observing that the P-star quark matter satisfies $n_d \simeq 2 \, n_u$, the quarks must hadronize into
neutron-rich nuclear matter. Moreover, in the hadronization processes we already estimated a gain in energy $E_{kin} \simeq 10$ MeV per
nucleon. Since:
\begin{equation}
\label{3.10}
E_{kin} \;  =  \;  \frac{m_N \, c^2}{\sqrt{ 1 - \beta_N^2}}  \; - \; m_N \, c^2 \; \; , 
\end{equation}
we readily obtain $v_N \simeq 0.15 \, c$. We end, thus, with the remarkable result that in the merger of two P-stars it is reasonably
to expect the launch of a relativistic hadron jet and unbound neutron-rich nuclear matter with  $M_{ej} \lesssim 10^{-1} M_{\bigodot}$
and average velocity  $v_N \simeq 0.15 \, c$.  It is worthwhile to mention that the final r-process abundance pattern is very robust
to variations in the detailed properties of the ejecta. This implies that the neutron-rich nuclear matter ejected in the collision of two
P-stars produces essentially final abundances identical to the neutron-star matter.
\section{Massive P-star binaries   }
\label{S4}
\begin{figure}
\vspace{-0.8 cm}
\begin{center}
\includegraphics[width=0.80\textwidth,clip]{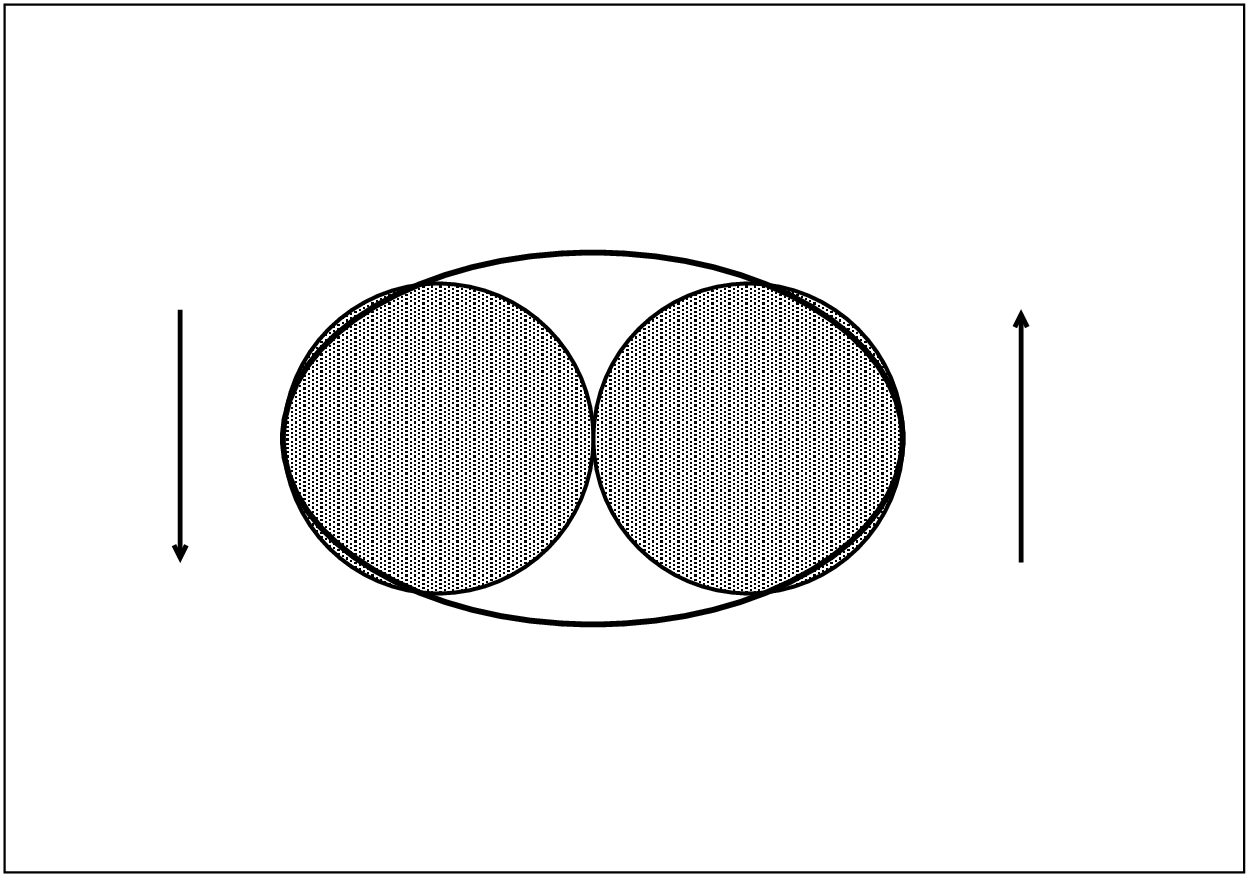}
\caption{\label{Fig6}
Schematic view of a massive P-star binary system in the post-merger phase. The arrows indicate the direction of rotation
around the rotational axis assumed to be perpendicular to the displayed plane.}
\end{center}
\end{figure}

The detection of GW150914 on 14 September 2015 by the two detectors of the Laser Interferometer Gravitational-wave 
Observatory~\cite{LIGO:2016a,LIGO:2016c} demonstrated how gravitational waves can be employed as a means to investigate
the properties of compact objects. In general,  compact objects are classified in accordance with  their mass.
According to widespread views,  if the compact object mass $M$ does not exceed the neutron-star maximum mass $M_{max}$,
then the compact star is classified as neutron star, otherwise it is identified with a black hole. There are been many efforts
to infer an upper bound on the maximum neutron-star mass. The detection of pulsars with masses above $ 2 \,  M_{\bigodot}$
suggests to set the limits on  $M_{max}$ in the range $2.3 - 2.5 \,  M_{\bigodot}$.  In the present paper we are following a more
conservative point of view by setting  $M_{max} \simeq 3.0 \,  M_{\bigodot}$. \\
Remarkably,  in almost all the gravitational wave events recorded in the first and second observing runs by the advanced
LIGO and Virgo observatories the masses of the compact binaries were well above  $M_{max}$, with the unique notable 
exception regarding GW170817 as discussed at length in Sect.~\ref{S3}. In particular, for the first gravitational wave event
GW150914 the  signal matched the waveform predicted by General Relativity for the inspiral and merger of a pair of black holes.
The initial black hole masses were $36^{+5}_{-4} M_{\bigodot}$ and $29^{+4}_{-3} M_{\bigodot}$, and the final black hole mass
was $62^{+4}_{-4} M_{\bigodot}$, with $3.0^{+0.5}_{-0.5} M_{\bigodot} c^2$ radiated in gravitational waves~\cite{LIGO:2016a,LIGO:2016c}.
As a matter of fact, this were the first observation of a binary stellar-mass black hole merger. \\
For an object with mass $M$ and spin $\vec{S}$, the dimensionless spin parameter is defined as:
\begin{equation}
\label{4.1}
\chi  \;  =  \;  \frac{c  \, |\vec{S}| }{G \, M^2}  \; \; \; . 
\end{equation}
It turned out that for GW150914 the spin of the more massive black hole was constrained by $\chi_1 < 0.7$, while the spin $\chi_2$
of the second black hole was only weakly constrained. On the other hand, the final black hole was a Kerr black hole with:
\begin{equation}
\label{4.2}
\chi _f \;  =  \;  a_f \; = \;  0.67^{+0.05}_{-0.07} \; \; , 
\end{equation}
where  $a_f$ is the dimensionless Kerr parameter. 
The merger of  compact binaries takes place in three stages. Initially, the compact objects circle around their common center of mass in
essentially circular orbits (the inspiral). They lose orbital energy in the form of gravitational radiation so that  they spiral inward.
In the second stage (the merging), the two objects coalesce to form a single final compact object. In the third stage (the ringdown),
the merged objects relaxes into its equilibrium state. \\
When the two bodies are far apart, their motion and waves can be described accurately by Newtonian gravity and the gravitational
radiation emitted during the inspiral stage can be evaluated by the simple Newtonian approach. At the end of the inspiral the binary
compact objects come crashing together and the orbits become very close and highly relativistic, just prior to radial plunge and binary merger.
Therefore, as the bodies spiral inward relativistic corrections become important. These relativistic corrections can be computed using a
post-Newtonian expansion of the Einstein field equations. Nevertheless, to compute the gravitational waves from the final merger
phase it is necessary to solve the Einstein field equations by means of numerical relativity (see, for instance, 
Refs.~\cite{Baumgarte:2010,Shibata:2016}). In the last stage of the binary merger the compact objects start accelerating radially inward,
plunging toward each other after they reach the so-called innermost stable circular orbit (ISCO) that for non-rotating objects
is given by:
\begin{equation}
\label{4.3}
r _{ISCO} \;  =  \;  6 \, \frac{G \, M}{c^2} 
\end{equation}
where $M$ is the total binary mass. The transition from inspiral to plunge is not sharp, so that the gravitational signal from the inspiral
phase is smoothly  connected to the ringdown gravitational signal. For black hole binaries the gravitational waves  by the
merging of the two black holes is produced through the so-called quasinormal emission that can be reliably estimated with the perturbation
theory in strong gravitational fields. During  the ringdown stage the merged object relaxes into its equilibrium state, that is a Kerr black hole.
In this case, the solution of the Einstein equations foreseen complex frequencies, the quasinormal modes, with the real part representing
the actual frequency of the oscillation and the imaginary part representing a damping. Thus, during the ringdown phase the graviatational
strain is basically a damped harmonic oscillator given by the lowest quasinormal mode. The damping time and ringing frequency of
this quasinormal mode depend only on the mass and dimensionless spin of the resulting Kerr black hole~\cite{Berti:2009}.
We see, then, that higher-order post-Newtonian calculations, together with significant contributions of numerical relativity and perturbation
theory in strong gravitational fields, have enabled the modelling of binary black hole mergers and rather accurate predictions of their
gravitational waveforms that have been successfully compared to observations. \\
We said that in the standard paradigm of Astrophysics a compact object with mass above $M_{max}$ is identified with a black hole.
However, we have seen that the  P-star equation of state allows for equilibrium stellar configurations with arbitrary masses.
Therefore, it is natural to address the question if some gravitational wave event can be accounted for by massive P-stars instead of
black holes. Unfortunately, in the case of P-stars there are not gravitational wave templates to be compared  with the observational data.
As a consequence, in the present Section  we must restrict ourself to, at best, a qualitative discussion. To be concrete, we shall attempt
a qualitative comparison to the first gravitational wave event GW150914. To this end, let us consider a binary system formed by
two non-rotating P-stars with approximatively equal mass:
\begin{equation}
\label{4.4}
M_1 \; \simeq \; M_2 \; \simeq \;   30 \,    M_{\bigodot} 
\end{equation}
and, according to Eq.~(\ref{2.25}):
\begin{equation}
\label{4.5}
R_1 \; \simeq \; R_2 \; \simeq \;  2 \sqrt{2} \; \frac{G \, M_1}{c^2} \;    \simeq \;  2 \sqrt{2} \; \frac{G \, M_2}{c^2} \; \; .
\end{equation}
It should be clear that the gravitational wave signal emitted by the coalescing P-star binary during the inspiral and merging phases is
practically indistinguishable from the case of two  Schwarzschild black holes with the same mass. In fact, P-stars have a finite tidal deformability
while for black holes the Love number vanishes. However, since massive  P-stars are the most compact stars consistent with causality their tidal
deformability is very small. Moreover, in general, the extraction of higher-order gravitational wave parameters, as the tidal deformability, from
the gravitational wave signal is difficult because these parameters can be efficiently extracted only in the late part of the inspiral, which
is not very long for massive binaries, and because there exist degeneracies between between different higher-order parameters, like
the spins of the compact objects. We are led, thus, to the noticeable fact that one can distinguish a massive P-star from a black hole
with the same mass only by looking at the gravitational waves emitted during the post-merger phase (the ringing phase for black holes).
 We will describe, now, the gravitational radiation emitted during the inspiral stage within the simple Newtonian approach. 
Gravitational waves are detected by measuring their effects on spacetime itself as the dimensionless strain $h = \Delta L/L$, where
$\Delta L$ is the fractional change of the length $L$. In General Relativity gravitational waves have only two independent
polarisation modes which, by convention, are referred   to as the plus and cross polarisation modes. Accordingly the strains from
these modes are $h_{+}$ and $h_{\times}$, respectively. Weak gravitational waves can be described by the ordinary plane wave solutions.
According to our qualitative discussion we shall assume:
\begin{equation}
\label{4.6}
 h_{+}(t) \;  \sim \;  h_{\times}(t) \;  \sim  \; h(t) \; \simeq \; h_0 \; \sin \Phi(t) \; \; \; .
\end{equation}
In the Newtonian approximation, for circular orbits,  the orbital frequency is given by:
\begin{equation}
\label{4.7}
 \omega_r  \;  = \;  \frac{G \, (M_1+M_2)}{R^3}   \; \; \; ,
\end{equation}
where $R$ is the orbital radius. The amplitude of the gravitational wave is readily obtained~\cite{Maggiore:2008}:
\begin{equation}
\label{4.8}
 h(t) \; \simeq \;   \frac{4}{r} \; \left ( \frac{G M_c}{c^2} \right )^{\frac{5}{3}}     \left ( \frac{\omega_{gw}}{2 \, c} \right )^{\frac{2}{3}}  
  \; \cos [ \omega_{gw} t \, + \, \Phi_0 ] 
\end{equation}
where
\begin{equation}
\label{4.9}
  \omega_{gw}   \;  = \; 2 \;  \omega_r  
\end{equation}
and
\begin{equation}
\label{4.10}
 M_c  \;  = \;  \frac{ (M_1+M_2)^{\frac{3}{5}}}{ (M_1 \, M_2)^{\frac{1}{5}}}  
\end{equation}
is the chirp mass.  Since the binary system loses energy due to the emission of gravitational waves, the orbital radius decreases leading
to an increase of $\omega_r$ according to Eq.~(\ref{4.7}). This, in turns, leads to an increase of  the power radiated in gravitational waves.
Then $R$ must decreases further leading to the coalescence of the binary system. Within our approximations one gets:
\begin{equation}
\label{4.11}
 \frac{d}{d \, t} \, \omega_{gw}(t) \; \simeq \;   2^{\frac{1}{3}} \; \frac{12}{5} \; 
 \left ( \frac{G M_c}{c^3} \right )^{\frac{5}{3}}     \omega_{gw}^{\frac{11}{3}}   \;  \; . 
\end{equation}
This last equation for the rate of change of the gravitational frequency is called the chirp. Integrating Eq.~(\ref{4.11}) we obtain:
\begin{equation}
\label{4.12}
 f_{gw}(\tau) \; \simeq \;   \frac{1}{\pi} \;  \left ( \frac{5}{256} \; \frac{1}{\tau} \right )^{\frac{3}{8}}    \;
  \left ( \frac{G M_c}{c^3} \right )^{ - \frac{5}{8}}   
\end{equation}
where $f_{gw} = \frac{\omega_{gw}}{2 \pi}$  and
\begin{equation}
\label{4.13}
\tau \; = \;  t _0 \; - \; t  \; \;  \; ,
\end{equation}
with $t_0$ the coalescing time.  Inserting Eq.~(\ref{4.12}) into Eq.~(\ref{4.8}) we get:
\begin{equation}
\label{4.14}
 h(t) \; \simeq \;   \frac{1}{r} \; \left ( \frac{G M_c}{c^2} \right )^{\frac{5}{4}}     \left ( \frac{5}{c \, \tau} \right )^{\frac{1}{4}}  
  \; \cos [  \Phi(\tau) ]  \; \; ,
\end{equation}
\begin{equation}
\label{4.15}
 \Phi(\tau)\; = \;   2  \pi \,    f_{gw}(\tau) \, \tau  \; + \; \Phi_0   \; \;  \; ,
\end{equation}
where $\Phi_0$ is an integration constant. Note that the gravitational wave frequency diverges when $\tau \rightarrow 0$. Actually, this 
divergence is never realised  since when the separation of the binary stars becomes smaller than $r_{ISCO}$
\begin{equation}
\label{4.16}
 r_{ISCO} \; = \; 6  \,  \frac{  G (M_1 + M_2)}{c^2}
\end{equation}
the two bodies merge (note that $r_{ISCO} > R_1 + R_2$). Taking into account that the binding energy for massive P-stars is almost
equal to the nucleon rest-mass energy (see Fig.~\ref{Fig2}), when the two P-stars coalesce they do not break apart but give rise
a structure with elongated shape that rotates around the center of mass with a certain rotational velocity as schematically illustrate
in Fig.~\ref{Fig6}.
After that,  the merged object, due to internal dissipative processes and the emission of gravitational waves, relaxes into an equilibrium
state that must be an almost spherical-symmetric P-star with final mass $M_f \simeq M_1 + M_2$ and radius $R_f > R_1$ since a 
P-star evolves by increasing the core energy density and by slightly expanding to fall into a stable branch of equilibrium configurations. \\
To estimate the gravitational waves emitted in the post-merger phase we attempt an exploratory study aimed to illustrate in the
clearest way the physics involved. To do this, we may approximate the newly born compact object to an ellipsoid with semi-axes
$a$, $b$ and $c$ that rotates around one of its principal axis (see Fig.~\ref{Fig6}). Within these approximations we can write:
\begin{equation}
\label{4.17}
a \; \simeq  \; c \; \simeq \; R_1 \; \; \; \; , \; \; \; \; b \; \simeq \; R_1 +  R_2 \; \simeq \; 2 \, R_1   \; \; ,
\end{equation}
so that the moments of inertia are:
\begin{equation}
\label{4.18}
I_1  \simeq  \frac{M_f}{5} (b^2 + c^2)  \simeq  M_f R_1^2  \; , \; 
I_2  \simeq   \frac{M_f}{5} (a^2 + c^2)   \simeq  \frac{2}{5}  M_f R_1^2  \; , \; 
I_3 \simeq   \frac{M_f}{5} (a^2 + b^2)   \simeq  M_f R_1^2  \; .
\end{equation}
Introducing the ellipticity:
\begin{equation}
\label{4.19}
\epsilon \; = \; \frac{I_1  \, - \, I_2}{I_3} \;  \sim  \; \frac{3}{5} \; \; , 
\end{equation}
we have for the gravitational wave amplitude~\cite{Maggiore:2008}:
\begin{equation}
\label{4.20}
 h(t) \; \simeq \;  \frac{ 4 \pi^2 G}{c^4}  \frac{I_3}{r}  \; (f^{'}_{gw})^2 \;  \; \epsilon \; 
 \cos \left ( 2 \pi f^{'}_{gw} \,  t \, + \, \Phi^{'}_0  \right )
\end{equation}
where $f^{'}_{gw} = \omega^{'}_{r}/\pi$. As a consequence the rotational energy of the body decreases due to the gravitational
wave emission:
\begin{equation}
\label{4.21}
\frac{d}{d \, t} \, E_{rot} \; \simeq \;   - \;  \frac{ 32 \,  G}{5 \, c^2} \; I_3^2    \; \epsilon^2 \; ( \omega^{'}_{r})^6 \; \; ,
\end{equation}
\begin{equation}
\label{4.22}
 E_{rot} \; \simeq \;  \frac{ 1}{2}  \; I_3   \;  ( \omega^{'}_{r})^2  \; \; \; .
\end{equation}
The ellipticity will rapidly decrease until it vanishes when the merged P-star reaches the final equilibrium configuration. Therefore,
we can write:
\begin{equation}
\label{4.23}
 \epsilon(t) \; \simeq \; \epsilon_0 \;  \exp  ( - \frac{t}{\tau_{dis}} )
\end{equation}
where the characteristic dissipation $\tau_{dis}$ is evidently of order:
\begin{equation}
\label{4.24}
\tau_{dis}  \; \sim \;  \frac{ R_1 + R_2}{c^{eff}_s}  \; \simeq   \;  2 \, \sqrt{2} \;  \frac{ G (M_1 + M_2)}{c^3}  \; \frac{c}{c^{eff}_s}     \; \; \; .
\end{equation}
In Eq.~(\ref{4.24}) $c^{eff}_s$ is an effective sound velocity that should satisfies $c^{eff}_s \lesssim c_s$. The inspiral phase of
the binary system lasts up to a time  $\bar{t} < t_0$, where $\bar{t}$ is roughly the instant time when the orbital radius is about $r_{ISCO}$.
In this phase the gravitational wave amplitude is given by Eq.~(\ref{4.14}). For $t > \bar{t}$ the binary P-stars plunge almost radially and finally
they merge. Since the gravitational signal from the inspiral phase is smoothly connected to the post-merger gravitational signal, we assume
that for  $t > \bar{t}$ the gravitational wave amplitude is given by:
\begin{equation}
\label{4.25}
 h(\tau^{'}) \; \simeq \;  \frac{ 4 \pi^2 G}{c^4}  \frac{I_3(\tau^{'})}{r}  \; [f^{'}_{gw}(\tau^{'})]^2 \;  \; \epsilon(\tau^{'}) \; 
 \cos \left ( 2 \pi f^{'}_{gw}(\tau^{'}) \,  \tau^{'} \, + \, \Phi^{'}_0  \right )
\end{equation}
where
\begin{equation}
\label{4.26}
\tau^{'} \; = \;  t  \; - \; \bar{t}  \; \;  \; .
\end{equation}
We may, also, safely neglect the time dependence of the moment of inertia. Accordingly, from Eqs.~(\ref{4.21}) and (\ref{4.22}) we get:
\begin{equation}
\label{4.27}
\frac{d}{d \, \tau^{'}} \, \omega^{'}_{r}(\tau^{'}) \; \simeq \;   - \;  \frac{ 32 \,  G}{5 \, c^2} \; I_3   \; [\epsilon(\tau^{'})]^2 \; 
[\omega^{'}_{r}(\tau^{'})]^5 \; \; .
\end{equation}
\begin{figure}
\centering
\includegraphics[width=0.48\textwidth,clip]{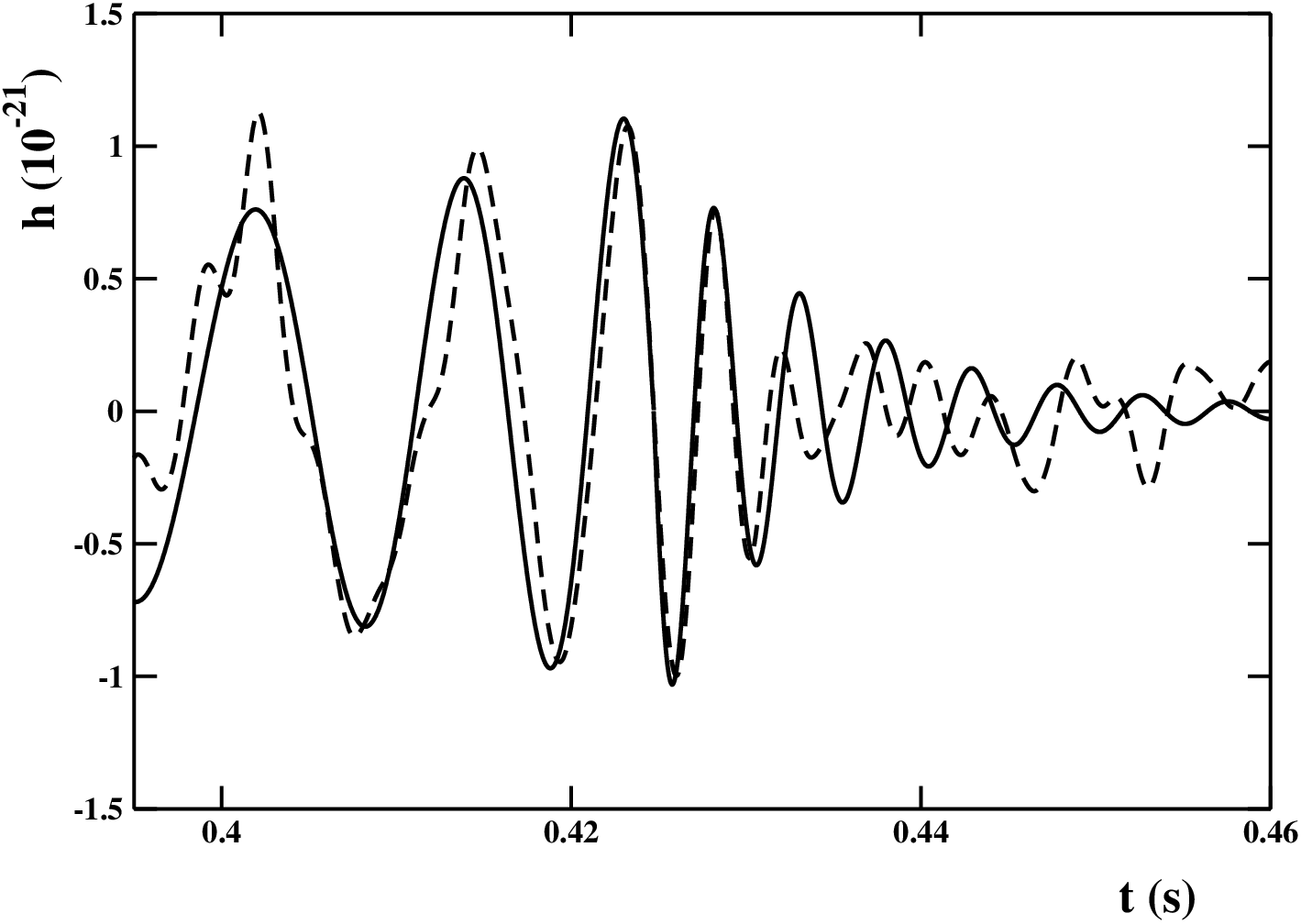}
\hspace{0.36 cm}
\includegraphics[width=0.48\textwidth,clip]{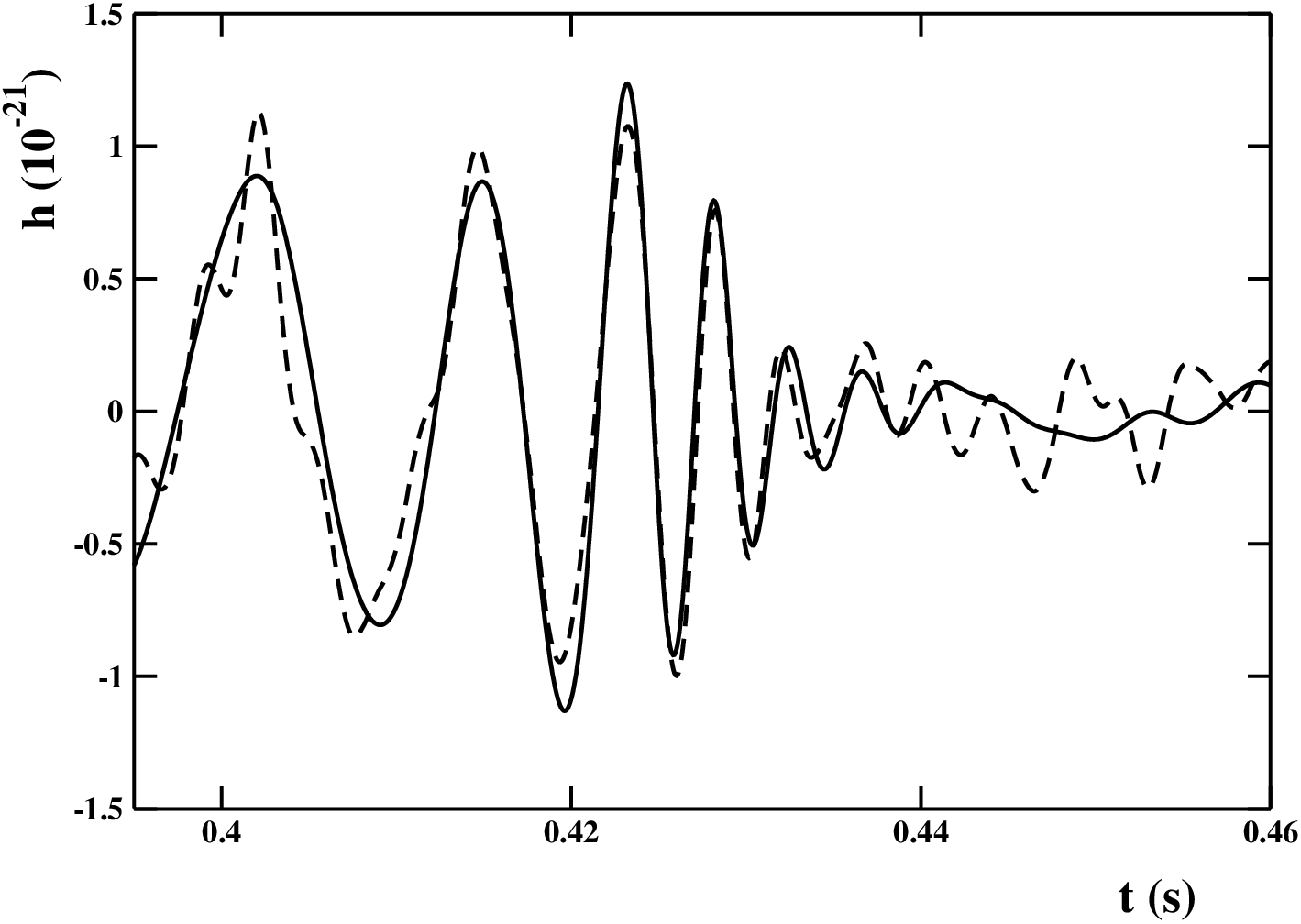}
\caption{\label{Fig7} 
Strain data for GW150914 as a function of time from the LIGO Handford detector~\cite{GW150914} (dashed line). The data from
the LIGO Livington detector are similar to those of the Handford detector.  The data have been filtered with a bandpass filter to suppress
large fluctuations and band-reject filters to remove instrumental spectral lines.
Comparison of the observed GW150914 strain data with the unfiltered P-star gravitational waveform, Eqs.~(\ref{4.14}) and
(\ref{4.25}) (left panel, full line) and the best-match template from non-rotating black-hole binary reconstructed using a
Bayesian analysis with the same filtering applied to the data~\cite{GW150914} (right panel, full line).}
\end{figure}
In Fig.~\ref{Fig7} we display the gravitational strain data for the gravitational event GW150914 from the LIGO Handford 
detector~\cite{GW150914}. From the data we infer:
\begin{equation}
\label{4.28}
t_0 \; \simeq  \;  0.4291131  \; \text{s} \;  \; .
\end{equation}
To compare the data with our estimate of the gravitational waves from our coalescing P-star binary we assumed:
\begin{equation}
\label{4.29}
\bar{t}  =  t_0  - \bar{\tau}   \; ,  \; \bar{\tau}  \simeq    0.0044 \, \text{s} \; , 
\;  \omega^{'}_{r}(\bar{t})  \simeq  1.5 \,   \omega_{r}(\bar{t})  \; ,  \;
\epsilon_0  \simeq  0.11  \; ,  \; \tau_{dis}  \simeq  4.0 \, 10^{-3}  \, \text{s}~\footnote{This corresponds to $c_s^{eff} \simeq 0.2 \, c$.}
\end{equation}
and fixed the phases $\Phi_0$ and $\Phi_0^{'}$ such that the gravitational strain from the inspiral phase is connected continuously to
the stain in the post-merger phase. In Fig.~\ref{Fig7}, left panel, we compare the gravitational wave for massive P-star binary as given
by Eqs.~(\ref{4.14}) and (\ref{4.25}) to the observed strain data for the gravitational event GW150914. We also show in Fig.~\ref{Fig7},
right panel, the best-match template from non-rotating black hole binary. We see that, notwithstanding our drastic simplifying assumptions,
the gravitational signal for coalescing massive P-stars in the post-merger phase seem to qualitatively  compare reasonably well with
the observational data. In other words,  the post-merger gravitational signal form massive P-stars could mimic the gravitational
wave emission from the final black hole during the ringdown stage. It is interesting to note that for the gravitational event  GW150914
the final Kerr black hole has a sizeable spin $a_f \simeq 0.67$. On the other hand, for P-stars  the remnant P-star has mass
$M_f \simeq 60   M_{\bigodot} $ and rather low rotational frequency $\nu_{f} \simeq 50$ Hz as given by the  asymptotic solution
of Eq.~(\ref{4.27}). \\
Finally, it is worthwhile to point out that a quantitative comparison with observations can be done only by means of gravitational
wave templates from the inspiral and merger of two massive P-stars obtained by solving the full Einstein equations numerically.
Nevertheless, it is interesting to note that for stellar-mass P-star binaries with highly asymmetric masses  the final merged star does 
not rotate, in general,
along a principal axis. As a consequence, the motion of the compact object is a combination of rotation around a principal axis and
precession. In this case, the gravitational waves emitted in the post-merger phase will contain not only the usual gravitational signal
at frequency $\omega_{gw} = 2 \omega_r$, but also a gravitational signal at the wobble frequency $\omega_{gw}/2= \omega_r$.
So that, the eventual detection of gravitational waves from asymmetric massive compact binaries in the post-merger phase 
with two main frequencies $\omega_{gw}$ and $\omega_{gw}/2$ would be a clear evidence of compact stars instead of black holes.
\section{Conclusions}
\label{S5}
Let us summarise the main results of the present paper. The purpose of this paper was to provide an adeguate discussion of our proposal
for a new class of compact relativistic stars, P-stars, composed by deconfined up and down quarks immersed in an almost
uniform chromomagnetic condensate. Our theoretical advance resided on our own understanding of the confining quantum vacuum
based on extensive non-perturbative studies of lattice quantum chromodynamics. The recent detection of gravitational waves from
merging binary compact systems together  with the observations of massive pulsars allowed us to critically compare our P-stars
with neutron stars. We argued that core-collapsed supernovae could give rise to a P-star instead of a neutron star. We showed that
the birth of  P-stars could completely solve the supernova explosion problem leading to both ordinary and superluminous supernova explosions.
We, also, found that P-stars are compatible with the gravitational wave event GW170817, that is widely believed to be the fingerprint
for neutron stars. Finally, we suggested that the gravitational waves from merging massive P-stars could mimic waveforms from
black hole binary coalescences in the ringdown phase. \\

To conclude, we feel that only observational evidences will be able to, ultimately, decipher the true nature of compact relativistic objects.

\end{document}